\documentclass[fleqn,usenatbib]{mnras}
\usepackage[T1]{fontenc}

\usepackage{xcolor}
\hypersetup{linkcolor=blue,citecolor=blue,filecolor=cyan,urlcolor=magenta}

\usepackage{hyperref}
\usepackage{graphicx}	% Including figure files
\usepackage{amsmath}	% Advanced maths commands
\usepackage{amssymb}	% Extra maths symbols
\usepackage{newtxtext,newtxmath}
\usepackage{footmisc}
\usepackage{deluxetable}
\def\mr{\mathrm}
\def\mc{\mathcal}

\def\d{\mr{d}}

\def\rin{r_{\rm in}}
\def\rout{r_{\rm out}}

\def\Msun{M_\odot}

\newcommand{\lrb}[1]{\left({#1}\right)}
\newcommand{\lrsb}[1]{\left[{#1}\right]}

\newcommand{\bs}[1]{{\boldsymbol{#1}}}

\def\hater{\hat{\bs{e}}_r}

\def\tj{t_{\rm on}}
\def\trec{P_{\rm prec}}
\def\Lj{L_{\rm j,iso}}
\def\rhoj{\rho_{\rm j}}
\def\rhow{\rho_{\rm w}}
\def\Gamj{\Gamma_{\rm j}}
\def\betaj{\beta_{\rm j}}
\def\Gamw{\Gamma_{\rm w}}
\def\betaw{\beta_{\rm w}}
\def\Gamh{\Gamma_{\rm h}}
\def\betah{\beta_{\rm h}}

\def\dotMw{\dot{M}_{\rm w}}
\def\tiletaj{{\eta}_{\rm j,iso}}
\def\etaj{\eta_{\rm j}}

\def\thej{\theta_{\rm j}}
\def\phij{\phi_{\rm j}}
\def\theLS{\theta_{\rm LS}}
\def\fb{f_{\rm b}}
\def\rg{r_{\rm g}}
\def\rd{r_{\rm d}}

\def\Omgp{\Omega_{\rm p}}
\def\rhoin{\rho_{\rm in}}
\def\pin{p_{\rm in}}

\def\rhowin{\rho_{\rm w,in}}
\def\pwin{p_{\rm w,in}}
\def\rhojin{\rho_{\rm j, in}}

\usepackage{ulem}

\title[precessing jets in TDEs]{Misaligned precessing jets are choked by the accretion disk wind}

\author[W. Lu et al.]{
Wenbin Lu$^{1}$\thanks{wenbinlu@berkeley.edu}, Tatsuya Matsumoto$^{2,3,4}$, Christopher D. Matzner$^{5}$\\
  $^1$Departments of Astronomy and Theoretical Astrophysics Center, UC Berkeley, Berkeley, CA 94720, USA\\
  $^2$Department of Astronomy, Kyoto University, Kitashirakawa-Oiwake-cho, Sakyo-ku, Kyoto, 606-8502, Japan\\
  $^3$Hakubi Center, Kyoto University, Yoshida-honmachi, Sakyo-ku, Kyoto, 606-8501, Japan\\
  $^4$Department of Physics and Columbia Astrophysics Laboratory, Columbia University, Pupin Hall, New York, NY 10027, USA\\
  $^5$Department of Astronomy and Astrophysics, University of Toronto, 50 St. George Street, Toronto, ON M5S 3H4, Canada
}

\begin{document}
\label{firstpage}
\pagerange{\pageref{firstpage}--\pageref{lastpage}}
\maketitle

\begin{abstract}

We analytically and numerically study the hydrodynamic propagation of a precessing jet in the context of tidal disruption events (TDEs) where the star's angular momentum is misaligned with the black hole spin.
% inside the accretion disk wind.
% This is expected the Lense-Thirring precession of a geometrically thick accretion disk with misaligned angular momentum around a spinning black hole.
We assume that a geometrically thick accretion disk undergoes Lense-Thirring precession around the black hole spin axis and that the jet is aligned with the instantaneous disk angular momentum. At large spin-orbit misalignment angles $\theLS$, the duty cycle along a given angle that the jet sweeps across is much smaller than unity. The faster jet and slower disk wind alternately fill a given angular region, which leads to strong shock interactions between the two. We show that precessing jets can only break out of the wind confinement if $\theLS$ is less than a few times the jet opening angle $\thej$. The very small event rate of observed jetted TDEs is then explained by the condition of \textit{double alignment}: both the stellar angular momentum and the observer's line of sight are nearly aligned with the black hole spin. In most TDEs with $\theLS\gg \thej$, the jets are initially choked by the disk wind and may only break out later when the disk eventually aligns itself with the spin axis due to the viscous damping of the precession. Such late-time jets may produce delayed radio rebrightening as seen in many optically selected TDEs.
% Our model leads to a number of predictions including delayed rebrightening of the radio afterglow emission and PeV neutrino emission from choked jets.
% We discuss a number of predictions from this model, one of which is the delayed radio rebrightening seen in many optically selected TDEs.
% \tm{One of the most important conclusions is that the very small event rate of jetted TDEs is explained by the condition of double alignment. I suggest to put it in the abstract.}
 
\end{abstract}


\section{Introduction}

Stars that are tidally disrupted by a supermassive black hole (BH) generally have misaligned orbits with respect to the BH spin axis. In most cases, it is expected that there is a substantial misalignment angle $\theLS\sim 1\rm\, rad$ between the orbital angular momentum ($\mathbf{L}$) of the accretion disk fed by the stellar debris and the spin angular momentum ($\mathbf{S}$) of a Kerr BH.
% A number of previous works \citep{stone12_jet_precession, franchini16_LT_precession_period, } have studied the evolution of disk warp and inclination --- affected by the Lense-Thirring torque from the BH, viscous coupling between different annuli, as well as the torque by the on-going fallback gas. 
At high accretion rates as expected in these tidal disruption events \citep[TDEs,][]{gezari21_TDE_review}, the disk is geometrically thick in the year or so \citep{strubbe09_TDE_disk_evolution} and strong viscous coupling enforces the entire disk to undergo Lense-Thirring precession around the BH spin axis like a solid body \citep[e.g.,][]{papaloizou95_warped_disk, fragile07_tilted_disk}. For an isolated accretion disk extending from the innermost stable circular orbit to roughly the tidal disruption radius, the precessional period is about 1 to 10 days for BHs near the maximum spin rate \citep{stone12_jet_precession, franchini16_LT_precession_period}. Recent works showed that the torque exerted by the on-going fallback gas may rapidly damp the disk precession and align the disk angular momentum with that of the original star provided that the dimensionless viscous parameter is sufficiently high $\alpha\gtrsim 0.1$ and the BH spin is significantly below maximum $\chi_{\rm bh}\lesssim 0.5$ \citep{zanazzi19_warped_disk_evolution} \citep[see also][]{xianggruess16_twisted_disk, ivanov18_twisted_disk_evolution}.

In this paper, we \textit{assume} that the disk undergoes solid body precession and explore the consequences of this hypothesis. In this picture, a relativistic jet launched near the vicinity of the BH will be collimated by the interactions with the wind from the outer disk and hence will precess around the BH spin axis \citep[see recent simulations by][]{liska18_precessing_jets}. As the jet axis precesses away from a given direction, the disk wind will refill that angular region. As a result, when the jet comes back to this direction, it must interact with the wind ahead of it. This interaction will be the focus of this paper.

We simplify the physical situation into the following model. On a spherical surface at a given radius (the inner boundary of our calculation), a relativistic jet is launched in the radial direction from a small solid angle (the ``jet cone''), the center of which precesses around the BH spin axis. At the same time, a slower disk wind is launched in the radial direction along the directions outside of the jet cone. We show that a narrowly beamed jet, with half opening angle $\thej\ll 1\rm\, rad$, with a large misalignment angle $\theLS\gg \thej$ is choked by the interaction with the disk wind, whereas a nearly aligned jet with $\theLS\lesssim \mbox{few}\,\thej$ successfully breaks out of the wind confinement. The latter case is in agreement with the results obtained by \citet{decolle12_jet_propagation}, who considered the propagation of a non-precessing jet aligned with the BH spin axis.

% We show that a narrowly beamed jet at a large inclination angle such that $\theLS\gg \thej$ is indeed choked by the interaction with the disk wind, whereas an aligned jet with $\theLS\lesssim \mbox{few}\times\thej$ successfully breaks out of the wind confinement. The latter case is in agreement with the results obtained by \citet{decolle12_jet_propagation}, who considered the propagation of a jet aligned with the BH spin axis.

% If we \textit{assume} that these precessing jets successfully escape from the system and produce bright X-ray emission, an observer whose viewing angle close to the misalignment angle $\theLS$ will see episodic emission with a duty cycle of the order $\thej/\theLS\ll 1$ in most cases \citep{stone12_jet_precession}, where $\thej$ is the half opening angle of the jet. 

A few TDEs with relativistic jets (hereafter ``jetted TDEs'') have been observed \citep[see][for a recent review]{decolle20_jetted_TDE_review}, including Swift J1644+57 \citep{bloom11_swj1644}, Swift J2058+05 \citep{cenko12_swj2058}, Swift J1112+82 \citep{brown15_swj1112}, and AT2022cmc \citep{andreoni22_AT2022cmc}. The rapid variability timescale of their X-ray emission indicates that the our line of sight is close to the jet axis and that the emission is strongly beamed towards the Earth. However, their X-ray lightcurves show an order-unity duty cycle, which has been used to argue\footnote{If we \textit{assume} that these precessing jets successfully escape from the system and produce bright X-ray emission, an observer whose viewing angle close to the misalignment angle $\theLS$ will see episodic emission with a duty cycle of the order $\thej/\theLS\ll 1$ in most cases \citep{stone12_jet_precession}.} that either the initial misalignment angle $\theLS$ is small \citep[$\lesssim$ few $\thej$,][]{stone12_jet_precession, lei13_J1644_misaligned} or the disk has already aligned itself with the BH spin by the time observations started \citep{tchekhovskoy14_J1644_MAD}. At sufficiently late time when the fallback rate drops to much below the peak value, we expect the disk angular momentum to gradually become aligned with the BH spin axis, as a result of viscous damping \citep{bardeen75_BP_alignment}, so the duty cycle of the X-ray emission from a persistent jet gradually approaches unity. However, the timescale over which the alignment occurs is much longer than the precessional period \citep[e.g.,][]{franchini16_LT_precession_period, zanazzi19_warped_disk_evolution}, and it is unlikely that the disk can align itself with the BH spin axis faster than the timescale for the peak fallback rate, which is roughly given by $t_{\rm peak}\sim 20\mr{\,d}\,(M/10^6M_\odot)^{1/2}$ ($M$ being the BH mass) for solar-like stars \citep[see][and refs therein]{coughlin22_peak_fallback}. At $t> t_{\rm peak}$, the mass fallback rate drops roughly as a power-law $\dot{M}_{\rm fb}\propto t^{-5/3}$ \citep{phinney89_fallback_rate, evans89_fallback_rate}. For the two jetted TDEs with nearly continuous X-ray monitoring, Swift J1644 and AT2022cmc, the power-law X-ray decay started within a timescale of two weeks or less after the initial discovery. This is inconsistent with viscous alignment of the disk and, instead, suggests that the angular momentum of the initial star was nearly aligned with the BH spin. The high peak isotropic X-ray luminosities ($\gtrsim 10^{47}\rm\, erg\,s^{-1}$) and large duty cycle factors of the other two jetted TDEs, Swift J2058+05 and Swift J1112+82, also suggest that the stellar angular momentum was nearly aligned with the BH spin.
% \tm{In this paragraph, would you like to discuss all observed jetted TDEs are actually aligned event?}
% depends on a number of highly uncertain aspects of the TDE system, but  

% : BH mass and spin, source(s) of viscosity, and the disk evolution including radiative cooling, winds and continuous fallback. Recent calculations by \citet{franchini16_LT_precession_period}, based on the \citet{shakura73_alpha_disk} model for the accretion disk, predict the alignment timescale for an isolated disk (ignoring the torques exerted by fallback material) to be months to years for a dimensionless viscosity parameter $\alpha\sim 0.01$ to 0.1. In reality, since the accretion disk is continuously fed by the fallback material (which generally has misaligned angular momentum), 

On the other hand, the volumetric rate of jetted TDEs for which our line of sight is within the relativistic beaming cone has been inferred to be $0.01$ to $0.1\rm\, Gpc^{-3}\,yr^{-1}$ based on the observed events \citep[e.g.,][]{andreoni22_AT2022cmc}. If we correct for an X-ray emission beaming factor of $f_{\rm b} = \thej^2/2 = 10^{-2}f_{\rm b,-2}$,
% that is inferred based on the isotropic equivalent of X-ray energy and the afterglow-informed jet kinetic energy \citep[e.g.,][]{decolle20_jetted_TDE_review},
the inferred intrinsic rate of jetted TDEs is in the range 1 to 10$\times f_{\rm b,-2}^{-1}\rm Gpc^{-3}\,yr^{-1}$. This is a very small fraction ($10^{-3}$ to $10^{-2}$) of the rate of all TDEs observed in various surveys, $\mc{R}_{\rm TDE}\sim 10^{3}\rm\, Gpc^{-3}\,yr^{-1}$ \citep[e.g.,][]{vanvelzen18_TDE_rate, sazonov21_erosita_TDEs, yao23_TDE_rate}. This is much smaller than the fraction of active galactic nuclei with jets, $\sim10\%$ \citep[e.g.,][]{Padovani+2017}. However, the small detection rate of jetted TDEs can be naturally explained, if the precessing jets in the majority of TDEs with large misalignment angles $\theLS\gg \thej$ are hydrodynamically choked 
% at early time when the fallback rate is high and
before the disk becomes aligned with the BH spin after a delay of months to years \citep[by which time the jet may shut off as the disk becomes radiatively efficient, e.g.,][]{shen14_TDE_disk_evolution}. In this picture, an observer will only detect bright X-ray emission from a successful jet at early time ($t\sim t_{\rm peak}$) under the \textit{double-alignment condition}: both the stellar angular momentum and the observer’s line of sight are nearly aligned with the black hole spin. One of the main goals of this paper is to quantify the probability of double-alignment condition (see eq. \ref{eq:double_alignment_probability}).

This paper is organized as follows. In \S \ref{sec:analytic_conditions}, we take a 1D approach to access the conditions for the breakout of a precessing jet, by considering an effectively episodic jet launched in a given direction. In \S \ref{sec:HD_sim}, we present 3D hydrodynamic simulations of a precessing jet propagating into a quasi-isotropic wind and compare the simulation results with our 1D analytic breakout conditions. Then in \S \ref{sec:discussion}, we discuss the potential caveats of our approach and future work that can address them. A summary of our results are provided in \S \ref{sec:summary}.

Near the completion of this work, we became aware of a paper by \citet{teboul23_precessing_jets} proposing a similar idea that misaligned precessing jets are choked by the disk wind. This work is complementary to theirs in two aspects: (i) our 1D analytic model is significantly different from theirs; (ii) we carried out 3D hydrodynamic simulations to test both analytic models. We compare our results to theirs in \S \ref{sec:HD_sim}.

\section{Analytic Model}\label{sec:analytic_conditions}
For an observer whose line of sight is at a polar angle $\theta\in (\theLS-\thej, \theLS+\thej)$ away from the BH spin axis, the jet power along the line of sight is episodic.  In this section, we analytically explore the conditions for successful jet breakout in the simplified picture of an episodic jet emitted into a steady wind, as an approximation to the case in which the jet precesses.
% and then apply the results to the case of a \citet{blandford77_BZjet} jet precessing at different inclination angles.
\subsection{An episodic jet in 1D}
We consider that, along a fixed direction, the jet is ``on'' for a duration of $t_{\rm on}$ and ``off'' for a duration of $P_{\rm prec}-t_{\rm on}$, where $P_{\rm prec}$ is the precessional period. The duty cycle is defined as
\begin{equation}\label{eq:duty_cycle}
    \xi_{\rm duty} = {t_{\rm on}\over P_{\rm prec}}\in (0, 1].
\end{equation}
When the jet is on, we assume for simplicity that it has a constant isotropic equivalent luminosity $\Lj$, bulk Lorentz factor $\Gamj\gg 1$ and speed $\betaj=\sqrt{1-1/\Gamj^2}$, so the density in the jet's comoving frame is given by
\begin{equation}\label{eq:jet_density}
    \rhoj' = {\Lj \over 4\pi r^2 \Gamj^2 \betaj c^3},
\end{equation}
where $r$ is the radial distance from the BH. Hereafter, primed $(')$ quantities are measured in the comoving frame of the fluid's bulk motion. When the jet is off, there is a steady baryonic wind with a constant mass-loss rate $\dotMw$, speed $\betaw$, and Lorentz factor $\Gamw=1/\sqrt{1-\betaw^2}$, so the density in the wind's comoving frame is
\begin{equation}\label{eq:wind_density}
    \rhow' = {\dotMw\over 4\pi r^2 \Gamw \betaw c}.
\end{equation}
Note that $\dotMw$ refers to the rate at which rest-mass is lost from the system whereas the relativistic mass of each particle is larger by a factor of $\Gamw$. In most physical cases, the wind is sub-relativistic, so we expect $\Gamw\approx 1$, but here we keep the relativistic expressions in this section (it is straightforward to take $\Gamw=1$ in the Newtonian limit).

We assume that the particles in the jet and wind are initially cold in the comoving frames of their bulk motion. The jet-wind interaction produces a forward shock which propagates into the wind and a reverse shock which propagates into the jet. We refer to the region sandwiched between the forward and reverse shocks as the ``jet head'' and denote its bulk Lorentz factor as $\Gamh$ and speed as $\betah$. If the inertia of the jet head is negligible, i.e. if changes in parameters near the forward shock produce nearly instantaneous changes near the reverse shock, then the two shocked regions that are separated by a contact discontinuity have the same pressure and velocity. We assume this to be true \citep[see][for a discussion when this assumption breaks down]{Uhm11_shock_jump_conditions}, and based on the Rankine-Hugoniot jump conditions at the two shocks and pressure balance, one obtains the velocity of the jet head \citep{matzner03_jet_head_propagation, matsumoto18_delayed_jet_breakout}
\begin{equation}\label{eq:jet_head_speed}
    \betah = \betaw + {\betaj - \betaw \over 1 + a^{1/2}\Gamw/\Gamj} = {\betaj + \betaw a^{1/2}\Gamw/\Gamj \over 1 + a^{1/2}\Gamw/\Gamj},
\end{equation}
where $a\equiv \rhow'/\rhoj'$ is the ratio between the two comoving densities. Note that $\betaw<\betah<\betaj$, meaning that the jet head always propagates slower than the jet behind it and faster than the wind ahead of it.
For an ultra-relativistic jet ($\Gamj\gg 1$), the lab-frame 4-velocities for the two shock fronts (fs$=$forward shock, rs$=$reverse shock) are given by the following approximations
% \tm{how did you derive them? (this is just my curiosity).}
\begin{equation}\label{eq:shock_4_velocity_approximate}
    u_{\rm fs} \simeq {4\over 3}\Gamh \betah, \ \ 
    u_{\rm rs} \simeq {3\over 2\sqrt{2}}\Gamh(\betah - 1/3).
\end{equation}
The above expressions show that the forward shock always propagates only slightly faster than the jet head, whereas the behavior of the reverse shock is more complicated: if the jet head is sufficiently slow $\betah\lesssim 1/3$ (meaning that the jet is running into a very dense wind), the reverse shock propagates backwards in radius; whereas if the jet head is sufficiently fast $\betah\approx 1$ (for a low-density wind), then the reverse shock propagates nearly as fast as the jet head.

% Using eqs. (\ref{eq:jet_density}, \ref{eq:wind_density}), we write
% \begin{equation}
%     {\Gamw a^{1/2}\over \Gamj} =  \lrb{{\Gamw\betaj\over \betaw} {\dotMw c^2\over \Lj}}^{1/2}.
% \end{equation}
It is convenient to define an isotropic equivalent jet efficiency factor, 
\begin{equation}\label{eq:jet_efficiency_factor}
% \boxed{
    \tiletaj \equiv {\Lj \over \dotMw c^2},
    % }
\end{equation}
and making use of eqs. (\ref{eq:jet_density}, \ref{eq:wind_density}, \ref{eq:jet_head_speed}), we write
\begin{equation}
    {\Gamw a^{1/2}\over \Gamj} = {\betaj - \betah \over \betah - \betaw} = \lrb{\Gamw\betaj\over \betaw\tiletaj}^{1/2}.
\end{equation}
Note that $\tiletaj$ is not the same as the conventional jet efficiency $\etaj$ --- the latter is defined as the physical jet power divided by the accretion rate onto the BH. In the \citet{blandford77_BZjet} framework, the conventional jet efficiency $\etaj$ is of the order $\mc{O}(\chi_{\rm bh}^2)$ for dimensionless black hole spin parameter $\chi_{\rm bh}$ and for the strongest possible magnetic fields in the black hole's magnetosphere \citep{tchekhovskoy11_jet_efficiency, narayan22_jet_efficiency}. We will discuss the physical values of $\tiletaj$ and $\betaw$ later based on observations and theoretical expectations, but for now, we stay agnostic to them. 

Let us first consider the race between the jet and the jet head. We see that the entire jet of radial thickness $\betaj c\tj$ will be shock-heated after a reverse-shock crossing time
\begin{equation}\label{eq:reverse_shock_crossing_time}
    t_{\rm cross} \simeq {\betaj\tj \over \betaj - \betah},
\end{equation}
where we have taken the velocity of the reverse shock to be roughly $\betah$. In fact, the reverse shock velocity $\beta_{\rm rs}$ is only close to $\betah$ when the jet head is highly relativistic (cf. eq. \ref{eq:shock_4_velocity_approximate}), and in this limit ($\betah \approx 1$), the reverse shock crossing time is reasonably accurate to within a factor of order unity. On the other hand, if the jet head is non-relativistic $\betah\approx 0$, the reverse shock speed is given by $\beta_{\rm rs} \simeq -1/3$, so we are only missing a factor of $4/3$ by taking $\betaj-\beta_{\rm rs}\simeq 1$ in this opposite limit. We also note that the reverse shock speed in eq. (\ref{eq:shock_4_velocity_approximate}) is obtained under the 1D picture where the shock-heated gas cannot exit the jet head region in the lateral direction. In reality, matter will exit the jet head region as long as $\thej \Gamh \lesssim 1$, in which case eq. (\ref{eq:reverse_shock_crossing_time}) is an even better approximation because the reverse shock will stay closer to the jet head.

We then consider the race between the jet head and the wind. We see that the jet head will catch up with the outer edge of the wind (which has radial thickness $\betaw c (\trec-\tj)$) after a breakout time
\begin{equation}
    t_{\rm bo} \simeq {\betaw(\trec - \tj) \over \betah-\betaw},
\end{equation}
where we have approximated the speed of the forward shock as $\betah$. In this picture, a successful jet breakout requires $t_{\rm bo}/t_{\rm cross} < 1$, which can be manipulated into the following form
\begin{equation}
    {t_{\rm bo} \over t_{\rm cross}} \simeq {\trec-\tj \over \tj} \cdot {\betaj-\betah \over \betah-\betaw} {\betaw\over \betaj} = {\trec-\tj \over \tj} \cdot \lrb{\Gamw \betaw \over \betaj \tiletaj}^{1/2} < 1.
\end{equation}
% where we have used eq. (\ref{eq:jet_head_speed_density_ratio}) for the ratio between the differential speeds.
% The above expression works for arbitrary
In this paper, we are interested in a special case of a relativistic jet ($\betaj\approx 1$) and a non-relativistic wind ($\betaw\ll 1$). In this case, we obtain the following simple criterion for successful jet breakout
\begin{equation}\label{eq:breakout_criterion}
% \boxed{
    {\xi_{\rm duty} \over 1-\xi_{\rm duty}} > \lrb{\betaw/\tiletaj}^{1/2}\ ,
    % }
\end{equation}
where $\xi_{\rm duty}$ (eq. \ref{eq:duty_cycle}) is the duty cycle of the episodic jet to be discussed in the next subsection.

\subsection{Duty cycle of a precessing jet}

\begin{figure}
\centering
\includegraphics[width=0.25\textwidth]{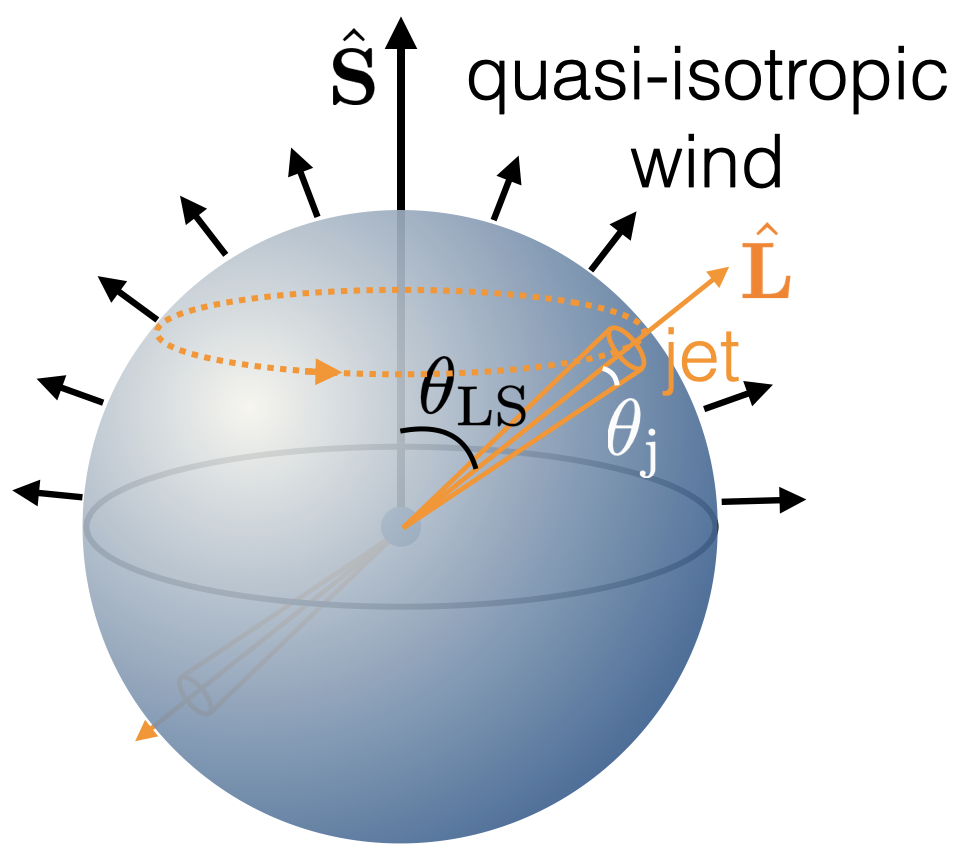}
\caption{Schematic picture of a highly misaligned precessing jet embedded in the disk wind. A narrow jet with half-opening angle $\thej$ whose axis precesses around the BH spin axis at an inclination angle $\theLS$. The jet is surrounded by a quasi-isotropic slower wind, which is launched in all directions except for the instantaneous jet cone. The hydrodynamic interactions between the jet and wind are the focus of this paper.
  }
\label{fig:geometry}
\end{figure}

Let us consider that a jet with half-opening angle $\thej\ll 1\rm\, rad$ that is precessing around the BH spin axis (hereafter the z-axis). The inclination angle between the jet axis and the z-axis is fixed at $\theLS \in [0, \pi/2)$. In reality, the jet has a non-trivial angular structure, but here we consider a ``top-hat'' jet for simplicity. A schematic picture of our consideration is shown in Fig. \ref{fig:geometry}.
% \tm{It would be helpful to put a schematic picture of Fig.~\ref{fig:geometry} here to clarify the geometry.}
% The precessional period is equal to the recurrence time $\trec$. 

\begin{figure}
\centering
\includegraphics[width=0.48\textwidth]{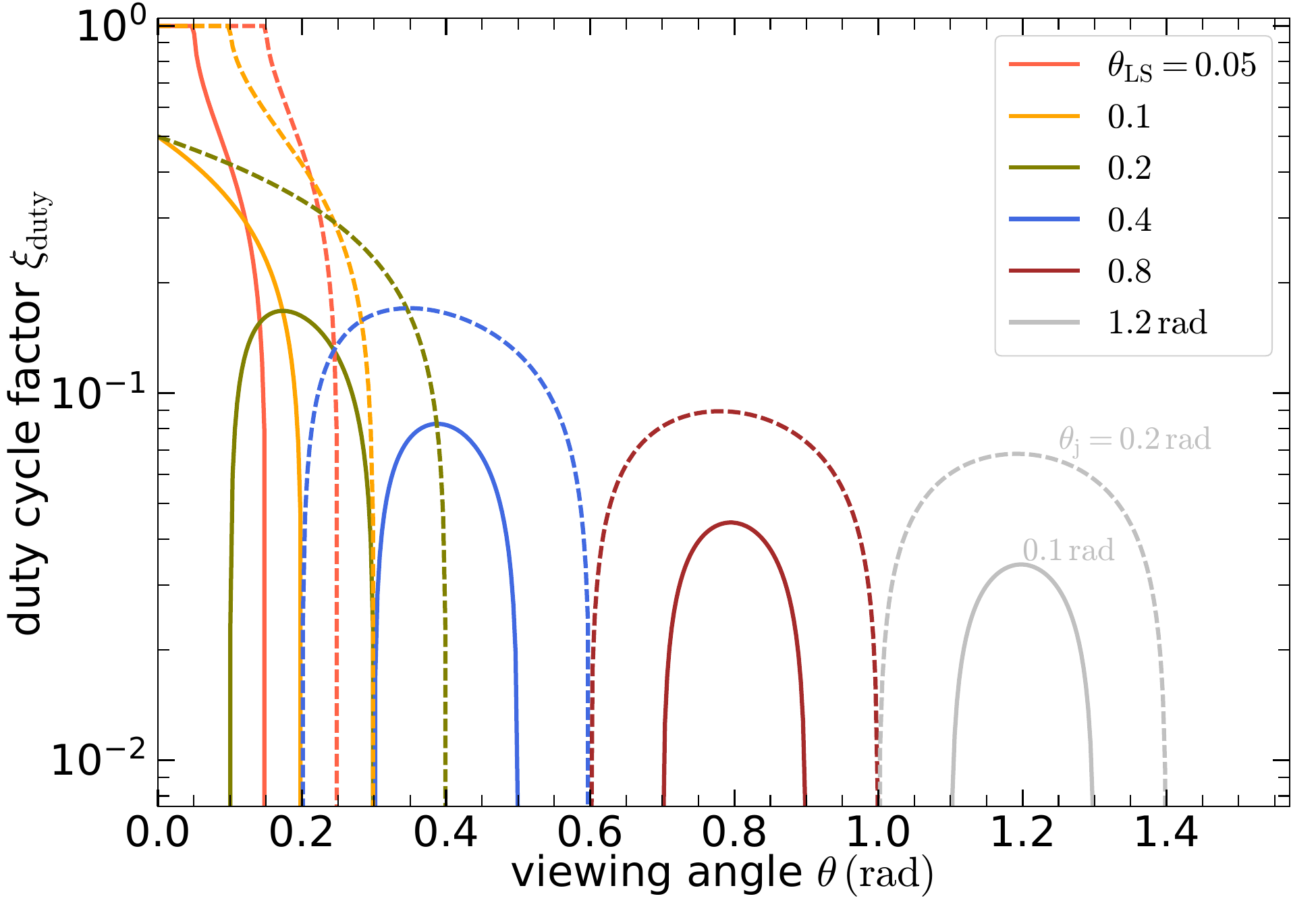}
\caption{Duty cycle for a ``top-hat'' precessing jet with different misalignment angles $\theLS$ and seen from different viewing angles $\theta$. The solid lines are for a narrow jet with an opening angle $\thej = 0.1\rm\, rad$ and the dashed lines are for a wider jet with $\thej=0.2\rm\, rad$. Different colors refer to a wide range of misalignment angles $\theLS$.
% We see that large duty cycle factors are only achieved in the ``double alignment'' cases when both the viewing angle and the misalignment angle are of the order the jet opening angle or less.
  }
\label{fig:duty_cycle}
\end{figure}

Consider an observer's line of sight at polar angle $\theta\in (0, \pi/2)$ and azimuthal angle $\phi=0$ in spherical coordinates. We restrict ourselves to $|\theta - \theLS|<\thej$ because otherwise the jet emission (which is assumed to be strongly beamed along the velocity vector $\vec{\beta}_{\rm j}$) will not reach the observer even if the jet breaks out successfully. The direction of the jet axis at a given time $t$ is specified by the polar angle $\theta=\theLS$ and azimuthal angle
\begin{equation}
    \phi_{\rm j}(t) = 2\pi t/\trec.
\end{equation}
The angle between the line of sight and the jet axis is denoted as $\Delta\theta$, the cosine of which is given by
% the following dot product
\begin{equation}
\begin{split}
    \cos\Delta \theta
    % &= \lrb{\sin\theta, 0,\cos\theta}\cdot \lrb{\sin\theLS \cos\phij, \sin\theLS \sin\phij, \cos\theLS}^{\rm T}\\
    % &=
    =\sin\theta \sin\theLS \cos\phij + \cos\theta \cos\theLS.
\end{split}
\end{equation}
The jet is only ``on'' when $\Delta \theta \leq \thej$, which gives a maximum azimuthal angle
\begin{equation}
    \cos\phi_{\rm j,max} = {\cos\thej - \cos\theta\cos\theLS\over \sin\theta \sin\theLS}.
\end{equation}
When the RHS of the above inequality is less than $-1$, which requires that both viewing angle $\theta$ and misalignment angle $\theLS$ are less than the jet opening angle $\thej$, the jet is always ``on'' and the duty cycle is $\xi_{\rm duty} = 1$. Such a jet will always break out from the wind.
% \footnote{Our consideration here does not include the case of pre-existing material, which may further prevent the jet from breaking out. \tm{I don't understand this footnote. What kind of pre-existing material are you thinking?}}.
Other than the cases with $\max(\theta, \theLS)<\thej$, the jet is only ``on'' along the observer's line of sight for at most part of the time (if at all). The duty cycle factors $\xi_{\rm duty}$ for different precessing jets are shown in Fig. \ref{fig:duty_cycle}. 
% Comparing the duty cycle factors with the jet breakout criterion in eq. (\ref{eq:breakout_criterion}), we see that the observer will only see strong jet emission when both the misalignment angle and the viewing angle are of the order the jet opening angle or less.

% \note{to be continued}

In the special case where the observer's line of sight is aligned with the jet axis with $\theta = \theLS$ (this ``on-axis'' case is best for detecting the jet emission), we obtain
\begin{equation}
    \cos\phi_{\rm j,max} = 1 - {1-\cos\thej\over \sin^2\theLS}\approx 1-{\thej^2/2\over \sin^2\theLS},
\end{equation}
where in the second expression we have taken the limit of $\thej\ll 1\rm\, rad$. For sufficiently large misalignment angles $\theLS> \thej$, we obtain $\phi_{\rm j,max}\approx \thej/\sin\theLS$, and hence the duty cycle is given by
\begin{equation}\label{eq:xi_on}
    \xi_{\rm duty, on\mbox{-}axis} = {\phi_{\rm j,max}\over \pi} \approx {\thej\over \pi \sin \theLS}, \ \mbox{ for } \theta=\theLS >\thej.
\end{equation}
From eq. (\ref{eq:breakout_criterion}), we obtain the maximum misalignment angle for a successful jet breakout
\begin{equation}
\begin{split}
    \theta_{\rm LS} \lesssim \theta_{\rm LS,max} &\simeq
    \frac{\theta_{\rm j}}{\pi}\left(\frac{L_{\rm j,iso}}{\dot{M}_{\rm w}\beta_{\rm w}c^2}\right)^{1/2}\\
 &\simeq 4\,\theta_{\rm j} \lrb{L_{\rm j,iso}/(10^{48}\mr{\,erg\,s^{-1}}) \over (\beta_{\rm w}/0.1) \, \dot{M}_{\rm w}/(\Msun \rm yr^{-1}) }^{1/2},
\end{split}
\end{equation}
where we have adopted the maximum duty cycle by $\xi_{\rm duty, on\mbox{-}axis}\simeq\theta_{\rm j}/(\pi\theta_{\rm LS})$.

The above model ignore the effects of the counter jet, which propagates in the opposite direction from the one we are considering here. The counter jet is only important for the largest misalignment angles and will change the duty cycle by a factor of 2 when $\pi/2-\theLS < \thej$.

\subsection{Applying the 1D model to the Blandford-Znajek jet}

In the framework of the \citet{blandford77_BZjet} jet model, we discuss the possible values of $\tiletaj$ (eq. \ref{eq:jet_efficiency_factor}) and $\betaw$. We start from the physical jet power, for which we conservatively assume that the accreting plasma near the BH is maximally magnetized and take the jet efficiency from recent simulations by \citet{narayan22_jet_efficiency}
\begin{equation}
    \etaj(\chi_{\rm bh}) \equiv {\Lj\fb \over \dot{M}_{\rm bh} c^2} \simeq 4\times10^{-3}\Phi_{\rm B}^2\Omega_{\rm H}^2 \lrb{1 + 1.38\Omega_{\rm H}^2 - 9.2\Omega_{\rm H}^4},
\end{equation}
where $\dot{M}_{\rm bh}$ is the accretion rate onto black hole, $\Omega_{\rm H}= \chi_{\rm bh}\rg /(2r_{\rm H})$ is the dimensionless angular frequency of the horizon, $r_{\rm H}/\rg = 1 + \sqrt{1-\chi_{\rm bh}^2}$ is the radius of the outer event horizon ($r_{\rm g}=GM_{\rm bh}/c^2$ is the BH's gravitational radius), and $\Phi_{\rm B}$ is the dimensionless magnetic flux threading the black hole's event horizon given by
\begin{equation}\label{eq:magnetic_flux}
    \Phi_{\rm B} \simeq -20.2\chi_{\rm bh}^3 - 14.9\chi_{\rm bh}^2 + 34\chi_{\rm bh} + 52.6,
\end{equation}
where $-1<\chi_{\rm bh} < 1$ is the BH spin parameter.
The wind mass-loss rate $\dotMw$ is generally not equal to the mass accretion rate onto the BH $\dot{M}_{\rm bh}$. In fact, the wind mass-loss rate might be much higher than the accretion rate onto the black hole by a factor of the order $(r_{\rm d}/r_{\rm g})^{s}$ and $0<s<1$ \citep{Blandford99_ADIOS}, where $r_{\rm d}$ is the radius of the outer disk that contributes to the majority of the wind mass-loss rate. In the following, we take $s=0.8$ \citep{yuan14_ADAF_review}, although \citet{begelman12_disk_wind_solution} argues for $s\approx 1$.
As for the beaming factor, we use $\fb = \thej^2/2$ for two jets beamed into two opposite cones each with a half opening angle of $\thej$. The wind speed is roughly given by $\betaw\simeq (r_{\rm g}/r_{\rm d})^{1/2}$ (i.e., the local Keplerian speed).

% The second, more important reason is that $\Lj$ is the isotropic equivalent jet luminosity, meaning that it is equal to the physical jet power divided by a beaming factor $f_{\rm b} = \thej^2/2$ if we assume two jets beamed into two opposite cones each with a half opening angle of $\thej$. For the physical jet power, we assume that 

Putting the above factors together, we obtain the RHS of the breakout condition in eq. (\ref{eq:breakout_criterion})
\begin{equation}\label{eq:breakout_condition_BZ}
    \lrb{\betaw/\tiletaj}^{1/2} \simeq {\thej (\rd/\rg)^{s/2-1/4} \over \sqrt{2\etaj(\chi_{\rm bh})}}.
\end{equation}
% As mentioned in earlier discussion, we expect the jet efficiency factor defined in eq. (\ref{eq:jet_efficiency_factor}) to be of the order $\tiletaj \sim \chi_{\rm bh}^2 \fb^{-1} \rg/\rd$ (for strongly magnetized plasma near the black hole), where $\chi_{\rm bh}$ is the dimensionless black hole spin parameter, $\fb=\thej^2/2$ is the jet beaming factor, and $\rd$ is the wind launching radius. The wind speed is roughly $\betaw\sim (\rg/\rd)^{1/2}$. Therefore, we may roughly estimate
% \begin{equation}
%     (\betaw/\tiletaj)^{1/2} \sim \chi_{\rm bh}^{-1} f_{\rm b}^{1/2} (r_{\rm d}/r_{\rm g})^{1/4} \simeq 0.1 \chi_{\rm bh}^{-1}  (\thej/0.1\mr{\, rad}) (r_{\rm d}/10r_{\rm g})^{1/4}.
% \end{equation}
It should be noted that for typical misalignment angles $\theLS\sim 1\rm\, rad$, the duty cycle factor scales linearly with the jet opening angle $\xi_{\rm duty}\propto \thej$ (eq.~\ref{eq:xi_on}). In the above equation, we also see that $(\betaw/\tiletaj)^{1/2}\propto \fb^{1/2} \propto \thej$. This means that, for large misalignment angles, the criterion for successful jet breakout mainly depends on the misalignment angle $\theLS$ and the black hole spin $\chi_{\rm bh}$ (and weakly on the wind launching radius $\rd$) but not on the jet opening angle $\thej$.

\begin{figure}
\centering
\includegraphics[width=0.48\textwidth]{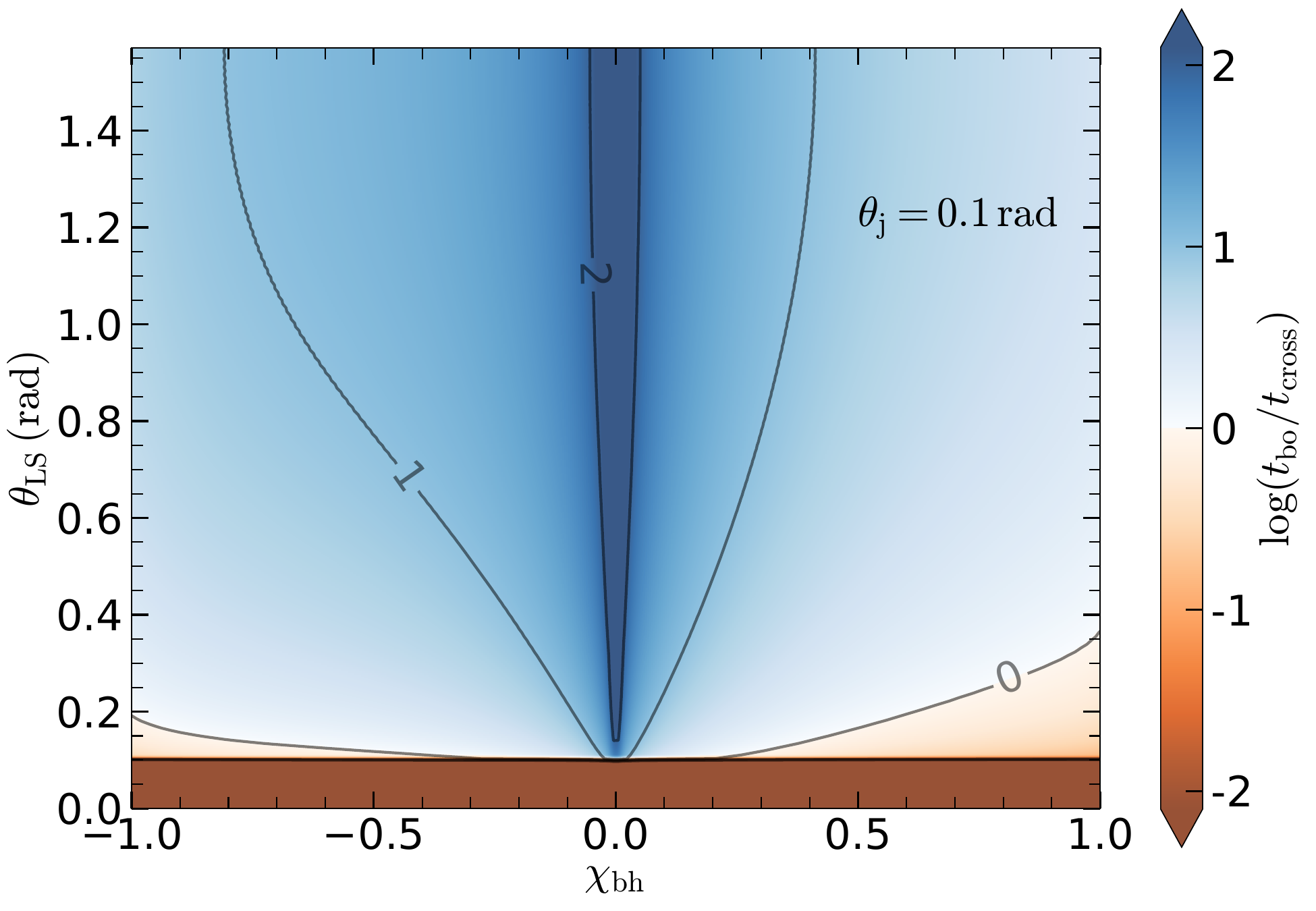}
\includegraphics[width=0.48\textwidth]{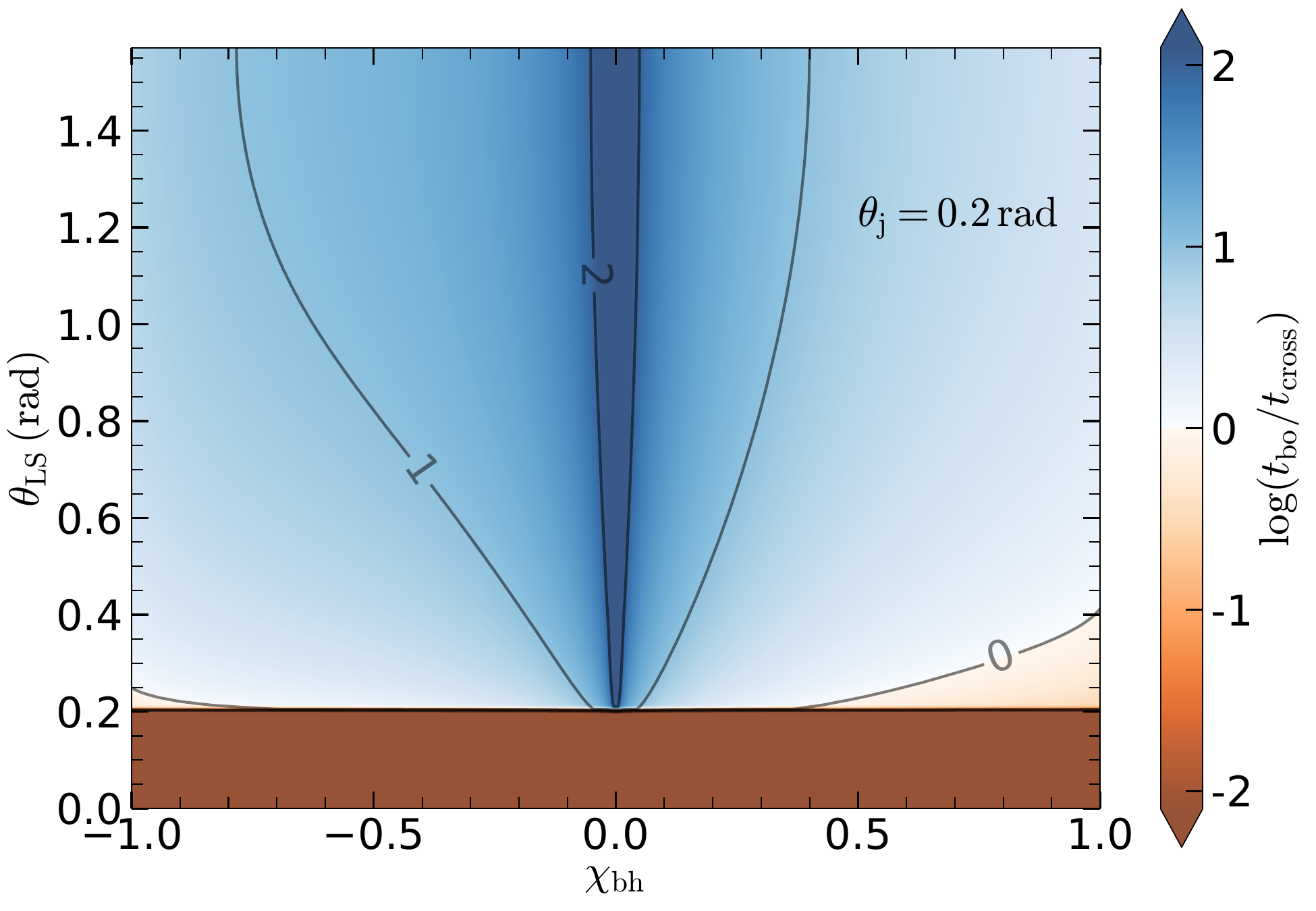}
\caption{The logarithm of the ratio between the breakout time and reverse-shock crossing time for the jet, $\log(t_{\rm bo}/t_{\rm cross})$, as a function of the BH spin $\chi_{\rm bh}$ and the misalignment angle $\theLS$. The jet power is based on the \citet{blandford77_BZjet} model. Successful breakout requires $\log(t_{\rm bo}/t_{\rm cross}) < 0$, as shown by the orange-colored region of the parameter space. The two panels are for different jet opening angles of $\thej=0.1$ (upper) and $0.2\rm\, rad$ (lower panel). For both panels, we fix the wind launching radius $\rd/\rg = 100$ (the results depend weakly on this parameter) and take the observer's viewing angle that maximize the duty cycle for a given misalignment angle $\theLS$ (i.e., $\theta\approx \theta_{\rm LS}$, the best case scenario for jet breakout). We note that our analytic breakout criterion (eq. \ref{eq:breakout_condition_BZ}) breaks down for aligned jets ($\theLS\lesssim\thej$) from the slowest spinning black holes $\chi_{\rm bh}\approx 0$, because in those cases, the pre-existing gas before the jet formation may be able to choke the extremely weak jets (such jets are difficult to observe anyway).
}
\label{fig:breakout}
\end{figure}

In Fig. \ref{fig:breakout}, we show the breakout condition $\log(t_{\rm bo}/t_{\rm cross})$ for a Blandford-Znajek jet and for a wind launching radius $\rd/\rg=100$,
% \tm{In the caption of Fig.~\ref{fig:breakout} you are saying $\rd/\rg=30$. Which is true? 30 seems more reasonable to TDEs.}
which is reasonable for typical TDEs of a solar-like star by a $10^6\Msun$ BH. 
% following Blandford-Znajek breakout condition as a function of the black hole spin $\chi_{\rm bh}$ and the misalignment angle $\theLS$,
% \begin{equation}\label{eq:breakout_condition}
%     \eta_{\rm bo,BZ} = \frac{\xi_{\rm duty} (\tilde{\eta}_{\rm j}/\beta_{\rm w})^{1/2}}{1-\xi_{\rm duty}}
%     \simeq {\xi_{\rm duty} \over 1-\xi_{\rm duty}} {\sqrt{2\etaj(\chi_{\rm bh})} \over \thej (\rd/\rg)^{1/4}},
% \end{equation}
% for two cases with $\thej = 0.1\rm\, rad$ and $\thej=0.2\rm\, rad$.
Successful jet breakout requires $t_{\rm bo}/t_{\rm cross} > 1$, which is only the case for small misalignment angles $\theLS\lesssim 0.4\rm\, rad$ (or $23^{\rm o}$) for all black hole spins. We conclude, based on our simplified 1D picture, that jets with large misalignment angles are always choked by the disk wind. In the next section, we present 3D hydrodynamic simulations of jet propagation and then compare the results with the 1D model.

\section{Relativistic Hydrodynamic Simulations}\label{sec:HD_sim}

In this section, we present 3D hydrodynamic simulations of a precessing jet interacting with the surrounding wind, as schematically shown in Fig. \ref{fig:geometry}.

\subsection{Numerical grid}

We use the relativistic hydrodynamics module of the $\mathtt{PLUTO}$ code \citep{mignone07_pluto_paper1, mignone12_pluto_paper2} and adopt a spherical grid $(r, \theta, \phi)$ for this problem. The radial grid is logarithmic with inner and outer boundaries $\rin$ and $\rout\gg \rin$. The jet and wind are injected at $\rin$ and we adopt an outflow boundary condition at $\rout$. The polar angle grid goes from $\theta=0$ to $\pi/2$ with linear spacing and $N_\theta=128$ points, and we adopt an outflow boundary condition at the equatorial plane\footnote{The more complete physical picture has two jets propagating in different directions. In this paper, we ignore the effects of the counter jet propagating away from our line of sight, because it only plays a minor role in the propagation of the forward unless the misalignment angle is close to $\theta_{\rm LS}\simeq\pi/2$.}. In the azimuthal direction, we adopt a uniform grid with $N_\phi=512$ in $[0, 2\pi)$ with periodic boundary conditions at both ends. Our grid roughly has angular resolution $\delta\theta\simeq \delta\phi\simeq 0.012\rm\, rad$, which is sufficient to resolve the angular structure of a jet with half-opening angle of $\thej\gtrsim 0.1\rm\, rad$. The outer radius is taken to be $\rout/\rin = 250$ and the number of grid points in the radial direction\footnote{The results presented here are from the high-resolution final runs, whereas our earlier attempts with lower spatial resolution by a factor of 1.5 in all dimensions produced similar results. This confirms the numerical convergence of our results.} is $N_r=448$, so the fractional radial resolution is $\delta r/r\simeq 0.012\rm\,rad$. We use the $\mathtt{TAUB}$ equation of state for adiabatic, perfect relativistic gas with an adiabatic index smoothly changing from $\gamma=5/3$ in the non-relativistic temperature limit to $4/3$ to the relativistic temperature limit. This choice of equation of state ignores the pressure contribution by radiation, which affects the compression ratio of the forward shock and should be addressed in future works.

\subsection{Boundary conditions and units}
Our machine units are such that
\begin{equation}
    \rin=1, \ \ c = 1,\ \ \rho_{\rm w,in}' = 1,
\end{equation}
where $\rho_{\rm w,in}'$ is the rest-mass density of the unperturbed wind in its comoving frame at the inner radius. In the physical cases of jetted TDEs, the jet is collimated by the wind from the outer disk which has a radius that is roughly equal to (at initial disk formation) or greater than (after some viscous evolution) the circularization radius of the bound stellar debris
\begin{equation}\label{eq:disk_radius_time_unit}
    \rd \gtrsim 2r_{\rm T} = 2R_*\lrb{M\over M_*}^{1/3}
    \simeq 1\mr{\, AU}\, {R_*\over R_\odot} \lrb{M/M_*\over 10^6}^{1/3}.
    % \rd/c \gtrsim 500\mr{\,s} {R_*\over R_\odot} \lrb{M/M_*\over 10^6}^{1/3}.
\end{equation}
In our numerical simulations, the inner radius of our grid $\rin$ roughly corresponds to the outer disk radius $\rd$ where the jet is collimated by the disk winds. For this reason, our time unit $t_{\rm in}=\rin/c$ can be considered to be of the order $10^3\rm\, s$ for typical jetted TDEs. In all expressions below, we keep the physical units for clarity.

The inner boundary conditions are described by a Gaussian-like isotropic equivalent kinetic power profile
\begin{equation}\label{eq:Lin}
    L_{\rm in}(\theta, \phi) = L_{\rm w} + (\Lj - L_{\rm w}) \exp(-\Delta \theta^2/\thej^2),
\end{equation}
a smooth 4-velocity profile\footnote{We choose a slightly wider opening angle of $1.5\thej$ for the 4-velocity profile to avoid too high jet densities near the edge of the jet, so the total mass-loss rate is dominated by the slow wind in all cases.}
\begin{equation}\label{eq:uin}
    u_{\rm in}(\theta, \phi) = u_{\rm w} + (u_{\rm j} - u_{\rm w}) \exp(-\Delta \theta^2/(1.5\thej)^2).
\end{equation}
% \begin{equation}
%     u_{\rm in}(\theta, \phi) = \max\lrsb{ u_{\rm j}\lrsb{1 + \lrb{\Delta \theta/\thej}^s}^{-k/s},\  u_{\rm w} }, \ \ k = 4, \ \ s=4,
% \end{equation}
% a Gaussian-like isotropic equivalent kinetic power profile
% \begin{equation}
%     L_{\rm in}(\theta, \phi) = \max\lrsb{L_{\rm j} \exp(-\Delta \theta^2/\thej^2),\  L_{\rm w}},
% \end{equation}
% a broken power-law isotropic equivalent kinetic power profile
% \begin{equation}
%     L_{\rm in}(\theta, \phi) = L_{\rm j}\lrsb{1 + \lrb{\Delta \theta/\thej}^s}^{-q/s} + L_{\rm w}, \ \ q = 5,
% \end{equation}
Here $L_{\rm w}$, $u_{\rm j}$, and $u_{\rm w}$ are the kinetic luminosity of wind, 4-velocity of jet and wind, respectively, which are discussed below.
In the above expressions, $\Delta\theta \in[0, \pi)$ is the angle between the radial unit vector $\hater(\theta,\phi)$ in the direction we are considering and the instantaneous geometrical center of the jet (i.e., the jet axis) $\hat{\bs{e}}_{\rm j}(\theLS, \phij)$, so we have
\begin{equation}
\begin{split}
    \Delta \theta(\theta, \phi, t) = \mr{acos} & [\sin\theLS\cos\phij \sin \theta \cos\phi\, +\, \\
    &\sin\theLS \sin\phij \sin\theta\sin\phi + \cos\theLS\cos\theta],
\end{split}
\end{equation}
where $\theLS$ is a fixed misalignment angle, $\phij(t)=\Omgp t$ is the azimuthal angle of the jet axis, and $\Omgp$ is angular frequency for the jet precession. Since the jet precessional period (days) is much shorter than the alignment timescale, we consider the misalignment angle to be constant in each simulation.

In our fiducial cases, we take a dimensionless precessional angular frequency of $\Omgp=0.03 c/\rin$, for which the precession period is 
\begin{equation}\label{eq:precession_period}
    P_{\rm prec}={2\pi\over \Omgp}=1.2\mr{\,d}\, {\rin \over \mr{\,AU}}\ ,
\end{equation}
which corresponds to a maximally rotationg BH, $\chi_{\rm bh}\simeq1$ \citep[e.g.,][]{franchini16_LT_precession_period}.
The longer the precessional period, the bigger the simulation box needs to be and the more computationally expensive the simulation is, because we must simulate a few precessional periods to reach a steady state. The outer boundary of the simulation is fixed at $\rout/\rin=250$, which corresponds to a physical outer radius of
\begin{equation}
    r_{\rm out} = 4\times10^{15}\mr{\,cm}\, {\rin\over \mr{\,AU}}.
\end{equation}
Our choice of $\rout=250\rin$ is much larger than the characteristic radius of the disk wind
\begin{equation}
    r_{\rm w}=\betaw c P_{\rm prec}\simeq 21\rin (\betaw/0.1) \simeq 3\times10^{14}\mr{\,cm}\, {\betaw\over 0.1} {\rin\over \mr{\,AU}}.
\end{equation}
Based on our analytic 1D consideration in \S \ref{sec:analytic_conditions}, if the jet head propagates beyond $r_{\rm w}$ before the reverse shock crosses the entire jet, the unshocked part of the jet will break out of the wind confinement. Our 3D simulations show that the jet continues to interact with the hot cocoon produced by earlier jet episode much beyond $r_{\rm w}$, and such continuing interactions are captured in our computational domain which spans about 10 times larger than $r_{\rm w}$. However, further interactions with the surrounding hot cocoon as well as the circum-nuclear medium do extend beyond our outer boundary $\rout$, which will eventually decelerate the jet and lead to its lateral expansion.

Observationally, the minimum variability timescale $t_{\rm var}\sim 10^2{\rm\, s}$ in the $\gamma$/X-ray lightcurves \citep{bloom11_swj1644, burrows11_Swift_J1644, yao23_AT2022cmc_X-ray} constrains the radius where the jet dissipates its energy and produces non-thermal emission
\begin{equation}
    r_{\rm diss}\sim c t_{\rm var}\Gamj^2 \simeq 3\times10^{15}\mr{\,cm}\, (t_{\rm var}/10^2\mr{\,s}) (\Gamj/30)^2,
\end{equation}
where we have taken a fiducial physical value\footnote{The jet Lorentz factor is constrained in the following (model-dependent) way. If the non-thermal $\gamma$/X-rays from the jet are produced by synchrotron emission, then the spectrum is expected to extend up to the ``synchrotron burnoff limit'' at photon energy of $\Gamj m_{\rm e} c^2/\alpha_{\rm FS} \sim 1\rm\, GeV (\Gamj/10)$ ($m_{\rm e}$ being the electron rest-mass and $\alpha_{\rm FS}$ being the fine-structure constant), which would seemingly overproduce the Fermi-LAT flux limits for Swift J1644+57 by 2 orders of magnitude \citep{bloom11_swj1644}. Thus, the GeV photons are most likely absorbed by the $\gamma\gamma\rightarrow e^\pm$ interactions with X-ray photons in the surrounding radiation bath, and the absorption optical depth $\tau_{\gamma\gamma}\gtrsim 1$ puts an upper limit to the jet Lorentz factor to be of the order $\mc{O}(10)$ for Swift J1644+57 and other jetted TDEs \citep{peng16_LAT_constraints}.} for the jet Lorentz factor $\Gamj = 30$ (although our current simulations cannot go to such a high Lorentz factor, unfortunately). On the other hand, the size of the thermal photosphere of the optically selected TDEs are of the order $10^{15}\rm\,cm$ \citep{gezari21_TDE_review}. Since our $\rout$ is comparable to the jet energy dissipation radius and the optical photospheric radius, it is likely that if the jet propagates up to $\rout$ without being significantly decelerated, it will produce bright non-thermal X-ray emission and hence be classified as a successful jet from an observational point of view. For these reasons, our outer radius $\rout$ should be sufficiently large to determine the success/failure of precessing jets. Future simulations should explore larger simulation boxes (more computationally expensive) and the electromagnetic emission from the (collisionless) shocks beyond $\sim 3\times10^{15}\rm\, cm$.
% \note{to be continued}

% but the jet emission could already be powered. variability time 100s. \note{to be continued}

\begin{figure}
\centering
\includegraphics[width=0.45\textwidth]{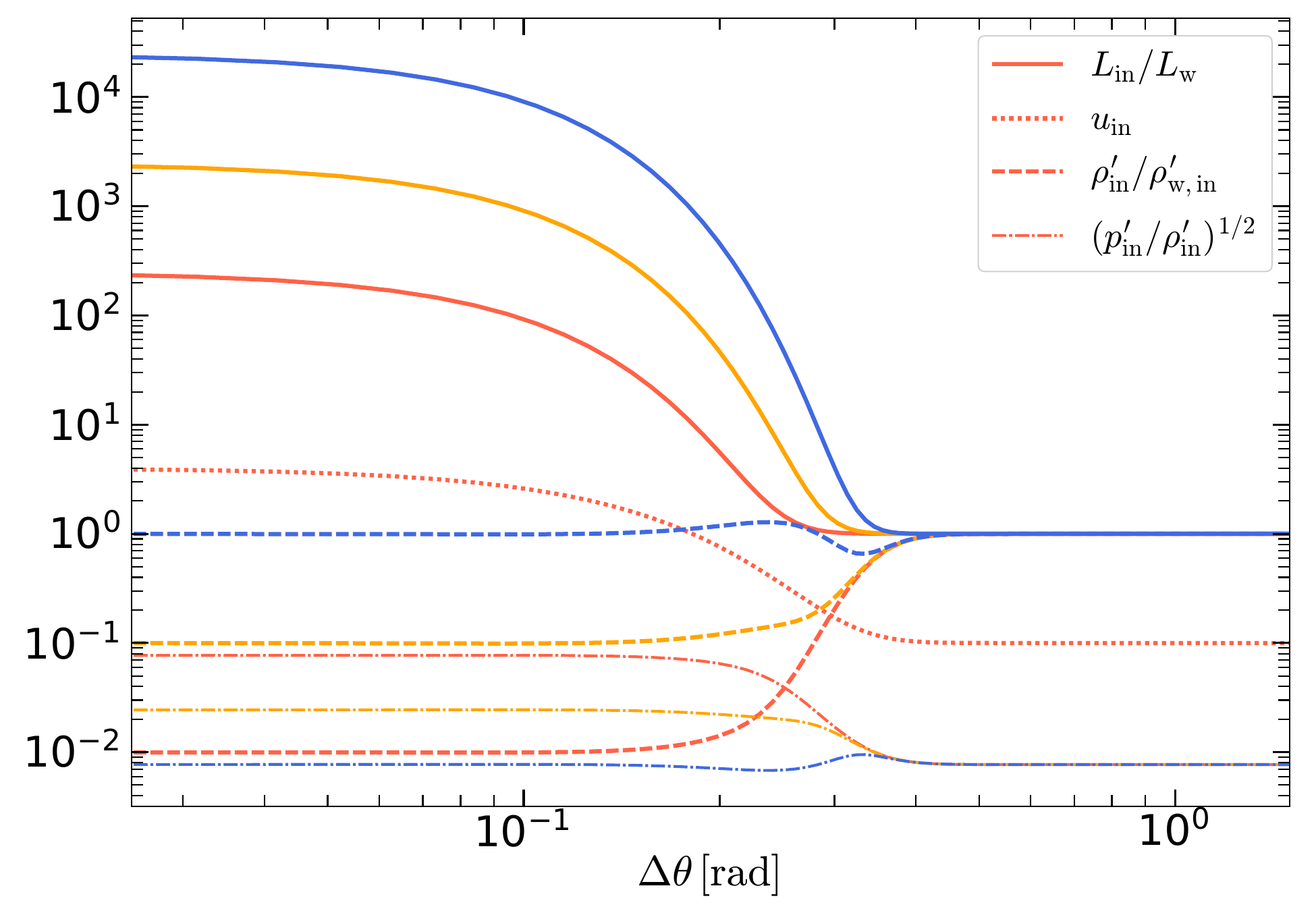}
\caption{Angular profiles for the jet+wind outflow adopted in our numerical simulations for the case with jet opening angle of $\thej=0.1\rm\, rad$. The three different colors are for $\rhojin'/\rhowin' = 0.01$ (red, weak jet), $0.1$ (orange, fiducial case), $1$ (blue, strong jet). The dash-dotted curves shows $(p_{\rm in}'/\rho_{\rm in}')^{1/2}$ which is roughly the comoving-frame sound speed of the gas at the inner boundary.
  }
\label{fig:jet_profiles}
\end{figure}

The wind 4-velocity at large angles from the jet axis is $u_{\rm w}$ and the maximum 4-velocity on the jet axis is $u_{\rm j}$. The system, even though already highly simplified, has a large number of free parameters, and we fix these two quantities as follows
\begin{equation}
    u_{\rm w} = 0.1, \ \ u_{\rm j} = 4.0.
\end{equation}
% which are consistent with our current understanding of jetted TDEs.
Hereafter, 4-velocities are in units of the speed of light $c$. The reason for the choice of a modest jet 4-velocity is that the causally connected region has a radial thickness of $\sim\! r/\Gamj^2\simeq 0.06r (\Gamj/4)^{-2}$ in the lab frame, so a higher jet Lorentz factor would require a much finer spatial resolution. 

We adopt a uniform comoving pressure profile in the lateral direction
\begin{equation}
    p_{\rm in}'(\theta, \phi) = \pwin' = {3\over 5\mc{M}_{\rm w}^2}\rhowin'\beta_{\rm w}^2 c^2,
\end{equation}
and the wind Mach number is taken to be $\mc{M}_{\rm w}=10.0$ for all cases, so the all fluid elements near the inner boundary are effectively presureless. 
The isotropic equivalent kinetic power $L_{\rm in}$ is related to the rest-mass density $\rhoin'(\theta, \phi)$ and pressure $\pin'(\theta, \phi)$ in the comoving frame by
\begin{equation}
    L_{\rm in}(\theta, \phi) = 4\pi \rin^2 c \lrsb{(\rhoin' c^2+\pin')\Gamma_{\rm in}u_{\rm in} - \rhoin' c^2 u_{\rm in}},
\end{equation}
where we have subtracted the rest-mass energy flux, and the Lorentz factor and velocity of the outflow at the inner boundary are given by $\Gamma_{\rm in}=\sqrt{1 + u_{\rm in}^2}$, $\beta_{\rm in} = u_{\rm in}/\Gamma_{\rm in}$. The isotropic equivalent kinetic powers of the outflows far from the jet axis (wind) and at the axis (jet) are given by
\begin{equation}
\begin{split}
    L_{\rm w} &= 4\pi r_{\rm in}^2c \lrsb{\rhowin' c^2 (\Gamw-1)u_{\rm w} + \pwin'\Gamw u_{\rm w}},\\
    L_{\rm j, iso} &= 4\pi r_{\rm in}^2c \lrsb{\rhojin' c^2 (\Gamj-1)u_{\rm j} + \pwin'\Gamj u_{\rm j}},
\end{split}
\end{equation}
so the 4-velocity and isotropic power profiles, $u_{\rm in}(\theta,\phi)$ and $L_{\rm in}(\theta,\phi)$, are given by eqs. (\ref{eq:Lin}, \ref{eq:uin}) at each coordinate time $t$. Then, the corresponding primitive gas variables at the inner boundary are
\begin{equation}
\begin{split}
    \rhoin'(\theta, \phi) &= {L_{\rm in}/(4\pi) - \pin'\Gamma_{\rm in} u_{\rm in} \over (\Gamma_{\rm in} - 1)u_{\rm in}}, \\
    \pin'(\theta, \phi) &= \pwin'.
\end{split}
\end{equation}
The jet angular profiles considered in this work are shown in Fig. \ref{fig:jet_profiles}.
% \tm{The description in this paragraph looks a little redundant to me. While you have already introduced the kinetic luminosity at the inner boundary by Eq.~(22) by using the wind and jet luminosities, it appears again in Eq.~(31) with pressure, density, and so on. This seems a little redundant, and I propose to remove Eq.~(31). But, at the same time I suspect my view is biased because I have never done simulations.}

\begin{table}
\centering
\begin{tabular}{c c c c c} 
 \hline
 N & $\rhojin'/\rhowin'$ & $\theLS/\rm rad$ & $\thej/\rm rad$ & $\Omgp/(\rin/c)$ \\
 \hline\hline
 4 & 0.1  & 0.1/0.2/0.4/0.8 & 0.1  & 0.03  \\
4 & 0.01 & 0.1/0.2/0.4/0.8 & 0.1  & 0.03  \\
4 & 1.0  & 0.1/0.2/0.4/0.8 & 0.1  & 0.03  \\
% 2 & 0.1  & 0.2/0.8 & 0.1  & 0.1  \\
3 & 0.1  & 0.2/0.4/0.8 & 0.2  & 0.03  \\
 \hline
\end{tabular}
\caption{A total number of $N=15$ simulations for various jet powers, misalignment angles, half-opening angles, and precessing rates have been performed. The following parameters have been fixed: $u_{\rm j} = 4.0$, $u_{\rm w}=0.1$, and wind Mach number $\mc{M}_{\rm w}=10.0$.}
\label{tab:sims}
\end{table}

Considering typical parameters for jetted TDEs (e.g., Swift J1644+57), the ratio between the jet and wind rest-mass densities is roughly given by
\begin{equation}
\begin{split}
    {\rhojin'\over \rhowin'} &\approx {\Lj/\Gamj^2\over \dotMw c^2/\betaw} \\
    &\simeq 0.11 {\Lj\over 10^{48}\mr{\,erg\,s^{-1}}} \lrb{\dotMw\over \Msun\mr{\,yr^{-1}}}^{-1}  \lrb{\Gamj\over 4}^{-2} {\betaw\over 0.1},
\end{split}
\end{equation}
or
% \tm{In some places you adopt $\dot{M}_{\rm w}=\Msun/yr$, but other places $\dot{M}_{\rm w}=2\Msun/yr$. Which one is our fiducial value?}
\begin{equation}
    \lrb{\tiletaj\over \betaw}^{1/2} = \lrb{\Lj \over \betaw \dotMw c^2}^{1/2} \approx 40\lrb{\rhojin'\over \rhowin'}^{1/2} {\Gamj/4 \over \betaw/0.1}.
\end{equation}

Our analytic arguments in \S \ref{sec:analytic_conditions} suggest that when $\sqrt{\tiletaj/\betaw}\gtrsim \xi_{\rm duty}^{-1}\sim \thej^{-1}$, the reverse-shock crossing time through the jet $t_{\rm cross}$ is longer than the jet head's breakout time $t_{\rm bo}$. 
% We consider two jet opening angles $\thej=0.1$ and $0.2\rm\, rad$
% from the wind launched between adjacent jet episodes.
% \note{Simulations indicate that the even after the jet breaks out of the local dense wind, the reverse shock continues to propagate into the jet such that the entire jet is eventually shock-heated and then laterally expands to a much wider angle than $\thej$.}
This motivates us to simulate three different jet densities
\begin{equation}
    \rhojin'/\rhowin' = 0.01, 0.1, 1,
\end{equation}
which determine the corresponding jet isotropic equivalent powers. Our simulations span four different misalignment angles on a logarithmic grid
\begin{equation}
    \theLS=0.1, 0.2, 0.4, 0.8\rm\, rad.
\end{equation}
This $3\times 4$ grid is sufficient to test our analytic arguments and inform us the true criteria for a successful jet breakout. Most simulations are for jet opening angle $\thej=0.1\rm\, rad$, but we also run a few cases with a wider opening angle $\thej=0.2\rm\, rad$ which gives a larger duty cycle factor. The parameters for all runs are summarized in Table \ref{tab:sims}.
% For these jet densities, the jet rest-mass energy density is always much higher than its pressure by at least a factor of 10.

% \note{to be continued}

% Going back to the physical situations of jetted TDE.

% Physically, perhaps $\rin$ corresponds to the outer disk radius at $10$--$10^2\rg$ where the jet is collimated, and the jet precessional period may be of the order one day \citep{franchini16_LT_precession_period}, so we have $\Omgp\sim 10^{-2}$--$10^{-1}$ in our machine units. However, from our analytic arguments, the jet breakout condition only depends on the duty cycle but not the precessional period, so we will test two choices of $\Omgp=0.03, 0.1$. 

% comment on rmax.

% To summarize, we have the following parameters (in machine units)
% \begin{equation}
% \begin{split}
%     u_{\rm j}&=4.0;\  u_{\rm w}=0.1;\  \pwin'=1.7\times10^{-2} \beta_{\rm w}^2; \\
%     \rhojin' &= 0.01, 0.1, 1; \  \Omgp = 0.03, 0.1;
% \end{split}
% \end{equation}

\subsection{Initial conditions}
% \note{to be continued}

We now turn to the initial conditions. Before the jet is launched, we consider that the simulation domain is pre-occupied by only the wind component with a uniform 4-velocity $u(r, \theta, \phi) = u_{\rm w}$ and an isotropic kinetic power of $L_{\rm w}$, from which we obtain the density profile and pressure profiles
\begin{equation}\label{eq:initial_condition_const_wind}
    \rhow'(r, \theta, \phi, t=0) = \rhowin' (r/\rin)^{-2},
\end{equation}
and the pressure profile follows adiabatic expansion of the wind $p_{\rm w}'(r, \theta, \phi, t=0) = \pwin' (r/\rin)^{-10/3}$. For numerical stability of the simulations, we also adopt density and pressure floor values of $\rho_{\rm floor}=10^{-6}\rhowin$ and $p_{\rm floor}=10^{-8}\rhowin c^2$, the latter of which avoids extremely large Mach numbers in the computational domain. These floor values are sufficiently low that they have little effects on the overall hydrodynamics of the system. Note that our initial conditions correspond to a long-lasting wind that has existed for on a timescale of 
\begin{equation}
    \rout/\betaw c \simeq 14\mr{\,d}\, (\rin/\mr{AU}) (0.1/\betaw),
\end{equation}
or longer before the jet launching. Such a situation may in fact be realistic the jetted TDEs because the hydrodynamical processes \citep[e.g., self-crossing of the fallback stream,][]{lu20_stream_self_intersection} prior to the disk/jet formation may fill up the space near the BH with dense gas. In future works, one might alternatively consider a situation that the wind kinetic power ramps up on some timescale, which will give a different density profile from that in eq. (\ref{eq:initial_condition_const_wind}).

\subsection{Results}

\begin{figure}
\centering
\includegraphics[width=0.48\textwidth]{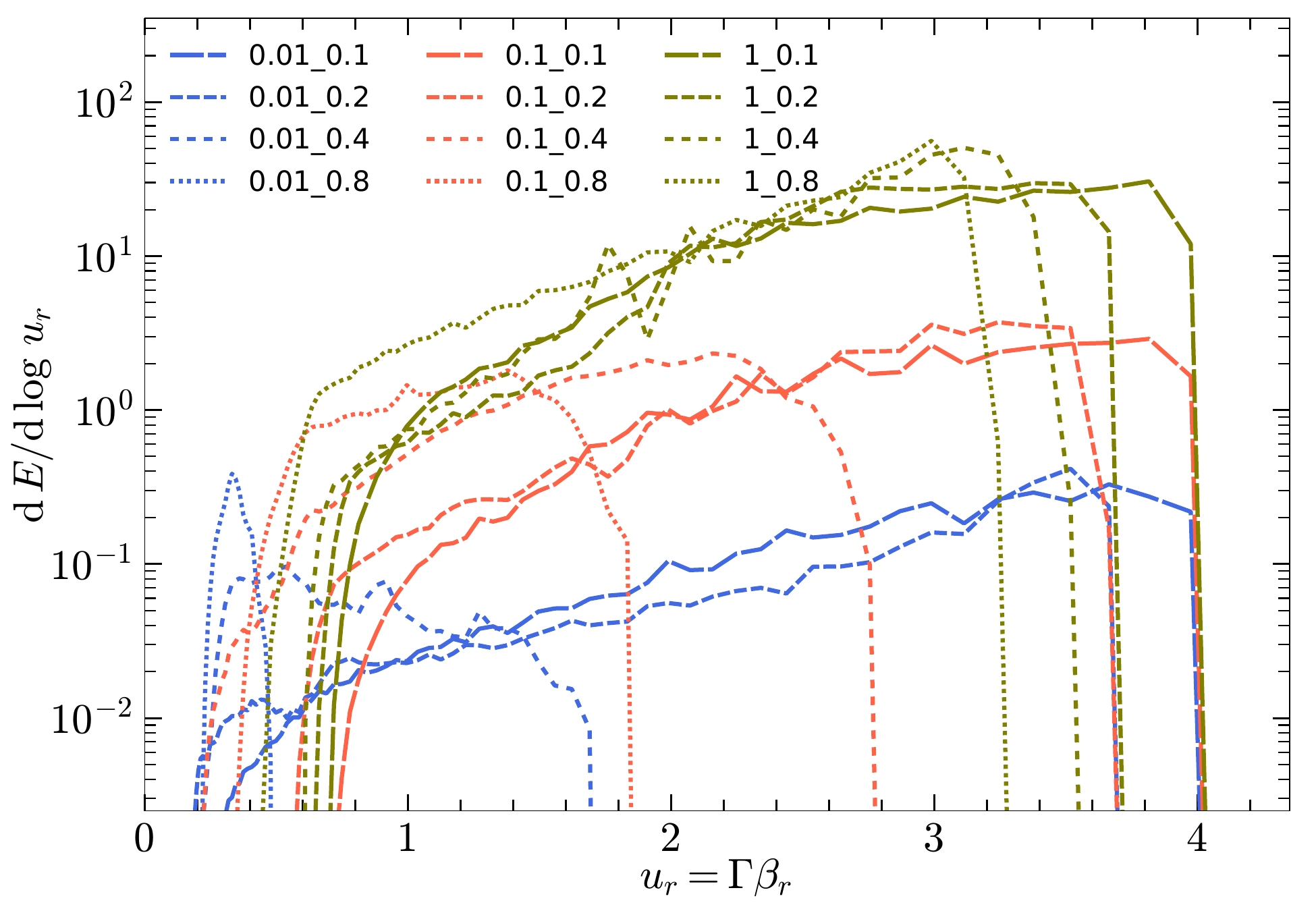}
\caption{Radial 4-velocity distribution for the gas in the radius range $200<r<250$ and between polar angles $\theLS-\thej$ and $\theLS+\thej$. Here, $\d E=(T^{00}-\gamma \rho' c^2)\d V$ is the kinetic energy of a fluid element of volume $\d V$, not including the rest-mass energy. The legend for each curve has two numbers separated by an underscore: the first one is $\rho_{\rm j,in}'/\rho_{\rm w,in}'$ ($\propto \Lj$) and the second one is the misalignment angle $\theLS$ in radian. We see that the peak of the radial 4-velocity distribution shifts to lower and lower values for increasing misalignment angles $\theLS$, especially for weaker jets with $\rho_{\rm j,in}'/\rho_{\rm w,in}'=0.01$ and $0.1$. This means that highly misaligned jets are more likely to be hydrodynamically choked by the disk wind.
% only the nearly aligned cases (0.1\_0.1, 0.1\_0.2, 1.0\_0.2) have successful jet breakout because the energy distribution peaks near $u_{\rm j}=4$.
  }
\label{fig:dEdur}
\end{figure}

\begin{figure*}
\centering
\includegraphics[width=0.75\textwidth]{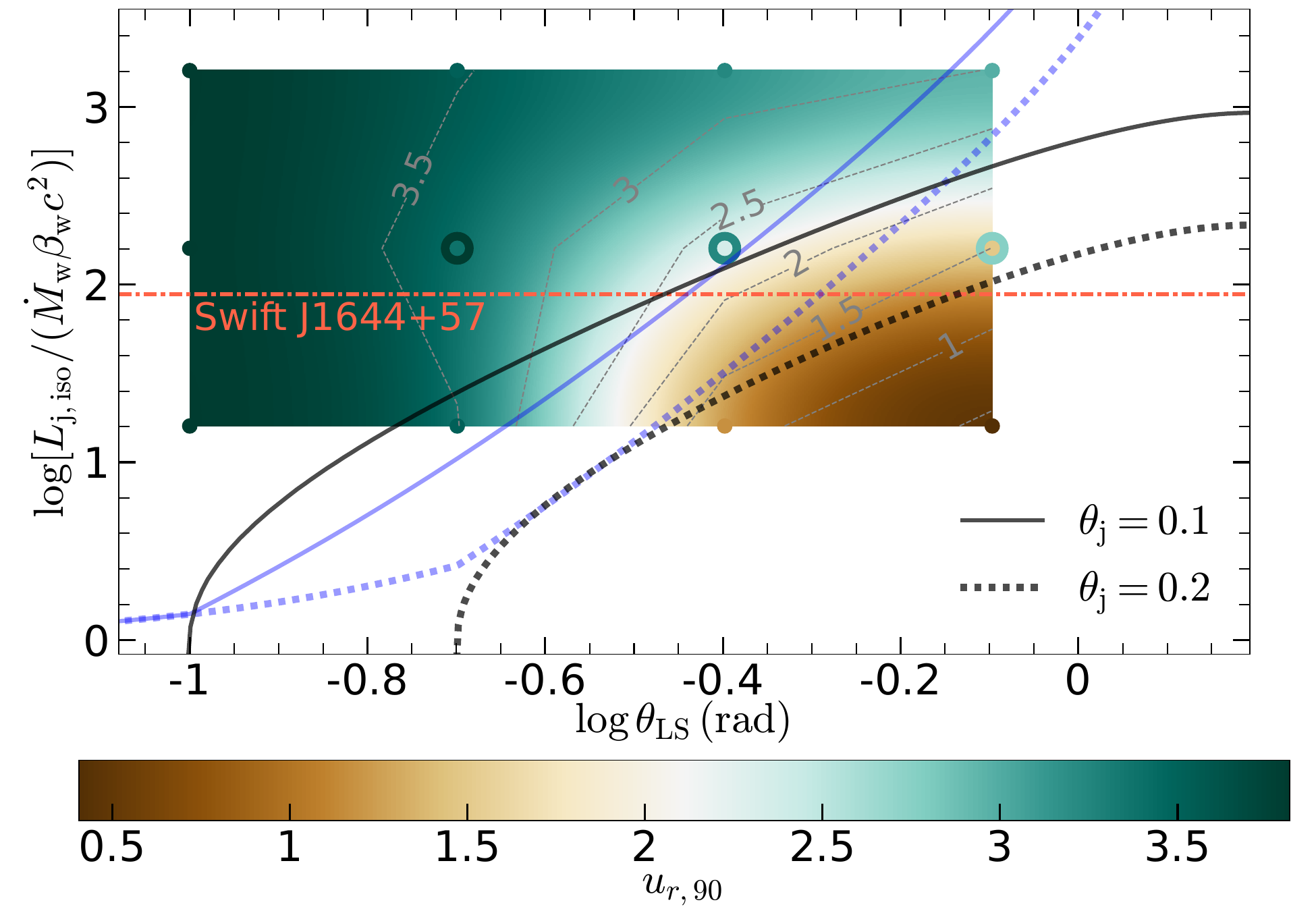}
\caption{Comparison between analytic jet breakout criteria (black/blue lines) and the results from numerical simulations (colored circles and image). The colors of the small circles show $u_{r,90}$ (eq. \ref{eq:ur90}) for a grid of simulations for $\thej=0.1\rm\, rad$, and the colored image is obtained from a bicubic interpolation. The large circles are for a wider jet with $\thej=0.2\rm\, rad$. The black solid and dotted lines shows our analytic criterion (eq. \ref{eq:breakout_criterion}) for the maximum duty cycle factor for two jet opening angles $\thej=0.1$ and $0.2\rm\,rad$, respectively. The blue solid and dotted lines show the analytic criterion (eq. \ref{eq:TM23}) proposed by \citet{teboul23_precessing_jets} for $\thej=0.1$ and $0.2\rm\,rad$, respectively. The horizontal red dash-dotted line shows a possible case like Swift J1644+57 but with different misalignment angles, with $\Lj=10^{48}\rm\, erg\,s^{-1}$, $\dotMw=2M_\odot\,\rm yr^{-1}$ and $\betaw=0.1$.
% \wl{[WL: I have added some contours to this figure.]}
  }
\label{fig:breakout_general}
\end{figure*}

We run each simulation until a steady state is reached and then analyze the velocity distribution. The final steady-state snapshots for three different jet luminosities/densities $\rhojin'/\rhowin' = 0.01, 0.1, 1.0$ and two different misalignment angles $\theLS=0.1, 0.8\rm\, rad$ are shown in Figs. (\ref{fig:rhojin0.1}, \ref{fig:rhojin0.01}, \ref{fig:rhojin1.0}) in the Appendix.

Fig. \ref{fig:dEdur} shows the distribution of kinetic energy in the radial 4-velocity space for the gas in the range of polar angles $\theLS-\thej < \theta < \theLS + \thej$ (the original jet beaming region) and in the radius range of $200 < r < 250$ (near the outer boundary). We find that the peak radial 4-velocity, where most kinetic energy is located, shifts to lower and lower values for increasing misalignment angles $\theLS$. This is in qualitative agreement with our analytic expectation that highly misaligned precessing jets are more likely to be choked by the disk wind. 

It is non-trivial to precisely define whether the jet in a given simulation partially breaks out of the wind confinement or is choked instead, mainly because of the difficulty of mapping the simulation results directly to observables.
% \tm{Observationaly, the size of photosphere of optical TDEs is $\sim10^{15}$cm, which is comparable to the outer boundary. So, if the jet arrives at this radius without significant deceleration, we would see a jet breaking through the outflow. This may reinforce the choice of our breakout criterion.}
Fortunately, we find that, when considering the radial 4-velocity distribution, the boundary between a partially successful jet and a choked jet is a reasonably sharp one.
% This is mainly due to the lack of understanding of the energy dissipation processes and X-ray emission mechanism\footnote{For instance, it is not clear if bright X-ray emission will still be produced when the maximum jet 4-velocity drops below $2$, which is significantly smaller than the initial value of $u_{\rm j}=4$ and it also means that the jet has undergone significant lateral expansion.}.
In this paper, we characterize the energy distribution shown in Fig. \ref{fig:dEdur} by $u_{r, 90}$ defined by
\begin{equation}\label{eq:ur90}
    P(u_{r}<u_{r, 90})=90\%,
\end{equation}
which means that 90\% of the kinetic energy in the $200<r<250$ and $\theLS-\thej < \theta < \theLS + \thej$ volume is carried by fluid elements with radial 4-velocity less than $u_{r,90}$. We also tried $u_{r,50}$ and $u_{r,80}$ by changing 90\% to 50\% and 80\%, and our qualitative conclusions are unchanged.

Fig. \ref{fig:breakout_general} shows $u_{r,90}$ as a function of misalignment angle and jet power for all our simulations. The small filled circles show the results from direct simulations for narrow jets with $\thej=0.1\rm\, rad$, and the colored image shows bicubic interpolation between the simulated grid points. Larger circles (only for 3 cases with $\rhojin'/\rhowin'=0.1$) are for wider jets with $\thej=0.2\rm\, rad$. The black solid line shows the analytic jet breakout criterion in eq. (\ref{eq:breakout_criterion}) for a narrow ``top-hat'' jet with $\thej=0.1\rm\, rad$, whereas the black dotted line is for $\thej=0.2\rm \, rad$.

We find that, for a given jet density $\rhojin'/\rhowin'$ or isotropic jet luminosity $\Lj$, there is a maximum misalignment angle $\theta_{\rm LS,max}$ beyond which $u_{r,90}$ is significantly reduced from $u_{\rm j}=4$ to a value around 2 or below. This maximum misalignment angle $\theta_{\rm LS,max}$ is roughly in agreement with our analytic breakout criterion (eq. \ref{eq:breakout_criterion}, black lines), except for the cases with the weakest jets $\rhojin'/\rhowin'=0.01$ for which our analytic result gives $\theta_{\rm LS,max}\simeq 10^{\rm o}$. It turns out that for very small misalignment angles $\theLS\lesssim 2\thej$, our 1D model (only considering the radial motion of the fluid elements) breaks down, because the hot cocoon surrounding the jet expands laterally to enshroud the entire region with polar angles $0<\theta\lesssim\theLS+\thej$. This makes it easier for the jet to break out from the confinement of the disk wind. In reality, it is possible that TDE jets have a more extended angular structure than the ``top-hat'' model considered in this work (e.g., \citealt{mimica15_J1644_afterglow, generozov17_afterglow_from_structured_jets}; but see \citealt{kumar13_swiftJ164457_radio} and \citealt{beniamini23_J1644_afterglow}), and that would also facilitate successful breakout for jets with small misalignment angles.

The blue lines in Fig. \ref{fig:breakout_general} show the jet breakout criterion proposed by \citeauthor{teboul23_precessing_jets} (\citeyear{{teboul23_precessing_jets}}, their eq. 42), which can be re-written in our notation as follows
\begin{equation}\label{eq:TM23}
    {\Lj \over \dotMw \betaw c^2} \gtrsim \max\lrb{1, {\theLS^2/\thej^2}} \lrb{1 + 4\pi^2\tan^2\theLS}.
\end{equation}
The blue solid solid line and dotted line are for $\thej=0.1$ and $0.2\rm\, rad$ respectively. The first factor of $\max(1, {\theLS^2/\thej^2})$ in the above expression comes from the argument that if the jet is indeed choked by the surrounding gas, then it will spread its energy over a much larger solid angle of the order $\pi\theLS^2$ than the the original jet solid angle of $\pi\thej^2$. The second factor of $1 + 4\pi^2\tan^2\theLS$ comes from a heuristic argument of a cylindrically helical motion of the jet head and then projection onto the z-axis ($=$ the BH spin axis). Our simulations show that the precessing jet drives a conic spiral (instead of cylindrically helix-shaped) forward shock into the disk wind, and the fastest moving fluid elements escape the system near the original cone of jet launching (at polar angle $\theta\sim \theLS$) instead of along the z-axis. This is likely the reason why the breakout criterion of \citet{teboul23_precessing_jets} does not capture the successful breakout of the most powerful jets at large misalignment angle (the case with $\rhojin'/\rhowin'=1.0$ and $\theLS=0.8\rm\, rad$ in our simulation). Nevertheless, for small misalignment angles $\theLS\lesssim 30^{\rm o}$, we find good agreements between our analytic breakout criterion and theirs, despite using very different physical arguments. 

If we adopt a critical value of $u_{r,90,\rm crit}\simeq 3$ below which the jet is considered to be choked, Fig. \ref{fig:breakout_general} shows that typical jetted TDEs, with $\Lj\simeq 10^{48}\rm\, erg\,s^{-1}$ (the peak $\gamma$/X-ray luminosity for observed jetted TDEs), $\dotMw\simeq 2M_\odot\rm\,yr^{-1}$ (the peak fallback rate), $\betaw \simeq 0.1$ (the Keplerian velocity for wind launching radius $\rd\simeq 100 \rg$), can only have a successful jet breakout provided that misalignment angle is sufficiently small $\theLS< \theta_{\rm LS,max} \simeq 15^{\rm o}$ for $\thej= 0.1\rm\, rad$. For a more conservative choice of $u_{r, 90, \rm crit}\simeq 2$, we find $\theta_{\rm LS,max}\simeq 20^{\rm o}$ for $\thej=0.1\rm\, rad$. For a larger jet opening angle $\thej=0.2\rm\, rad$ but keeping $\Lj$ and $\dotMw$ fixed, it is much easier for the jet to break out of the wind confinement (see the results shown by large circles in Fig. \ref{fig:breakout_general}); however, if we keep the ratio between the physical (beaming-corrected) jet power $\Lj \thej^2/2$ and the wind mass-loss rate $\dotMw$ fixed, then the maximum misalignment angle for a successful jet breakout should remain around $\theta_{\rm LS,max} \sim 15 \mbox{ to } 20^{\rm o}$.
% Based on our model, for the vast majority of the jetted TDEs, the misalignment angles are so large that the jets are choked (at least at early time when the fallback rate is near the peak value).

% \note{to be continued}

\section{Discussion}\label{sec:discussion}

In this section, we discuss potential caveats in our model and then suggest directions for future research.

(1) We only consider hydrodynamic jets whereas magnetic fields may play important roles in jet launching as well as propagation \citep[e.g.,][]{tchekhovskoy15_jet_launching}. A Poynting-dominated jet is subjected to MHD instabilities (e.g., the kink instability), which can lead to bending of the jet and energy dissipation that are not captured in our simulations \citep[e.g.,][]{bromberg16_MHDjets, barniolduran17_MHDjets}. Our qualitative conclusion --- highly misaligned precessing jets are choked by the disk wind, should remain unchanged as any jet must do a similar amount of mechanical work to break out of the confinement of the disk wind. Recent GRMHD simulations of misaligned accretion disks by \citet{liska23_precessing_jet_GRMHD} show some hint that weak precessing jets may be choked by the disk wind (see their Fig. 2), but their simulation domain only captures the interactions on small scales ($\lesssim \rm\, AU$ for a $10^6M_\odot$ BH). Future works should study to propagation of Poynting-dominated precessing jets on larger scales (1 to $\gtrsim10^2\rm\, AU$) as done in this paper. 

(2) We have assumed that the disk wind is quasi-isotropic whereas in reality the mass-loss rate and speed of the wind are functions of the angle $\Delta \theta$ away from the (instantaneous) jet axis \citep[e.g.,][]{thomsen22_superEdd_outflows}. For realistic the angular dependence of $\dotMw(\Delta \theta)$ (the isotropic equivalent wind mass-loss rate) and $\betaw(\Delta \theta)$, our simulation parameters should correspond to the functional values at $\Delta \theta\simeq \theLS$.
% because the jet-wind interactions mainly occur at large angles $\Delta \theta\sim \theLS$ away from the jet axis where these two angle-dependent functions are not expected to vary strongly across an angular size of the jet opening angle $\thej$. 
Future works should include the full functional forms of $\dotMw(\Delta \theta)$ and $\betaw(\Delta \theta)$.

(3) The pressure in the disk wind is dominated by radiation, which affects the equation of state of matter in our simulation. The electron-scattering optical depth of the wind near the characteristic wind radius $r_{\rm w}=\betaw c P_{\rm prec}$ (where most parts of the jet-wind interactions occur) is given by
\begin{equation}\label{eq:wind_optical_depth}
    \tau(r_{\rm w}) \simeq {\dotMw\kappa_{\rm s}\over 4\pi r_{\rm w} \betaw c} \simeq 2 {\dotMw\over \Msun\rm\,yr^{-1}} \lrb{\betaw\over 0.1}^{-2} {\mr{\,AU}\over \rin},
\end{equation}
where we have taken $\kappa_{\rm s}=0.34\rm\, cm^2\,g^{-1}$ for solar metallicity and the precessional period given by eq. (\ref{eq:precession_period}).
% \tm{This estimates implies that the photosphere of optical TDEs is around $\sim10^{14}\,\rm cm$, a little smaller than observed values? But please ignore this comment if you never discuss the relation between wind and optical emissions.}
We see that typical jetted TDEs have $\tau(r_{\rm w})\lesssim \betaw^{-1}$, which means that radiation in the unshocked wind at radii $r\sim r_{\rm w}$ can quickly diffuse away\footnote{The relatively fast ($\sim\! 0.1c$) disk wind considered in this work is unlikely produce the bright optical emission seen in many TDEs, because the optical photospheric radii are of the order $10^{15}\rm\, cm$. Interestingly, eq. (\ref{eq:wind_optical_depth}) suggests that we may expect the jet-wind interactions considered in this paper to produce a shock-breakout-like signal in the extreme-UV and soft X-ray bands.}. But even so, the radiation pressure still dominates over the gas pressure for typical TDE parameters. We see that radiation-hydrodynamic simulations are needed to correctly capture the equation of state of matter. This may weakly change the results presented in this paper (by affecting the speed of the jet head $\betah$) and should be explored in future works. 
% The pressure of the shock-heated wind is roughly given by
% \begin{equation}
%     p_{\rm sh} \simeq {\Lj\over 4\pi r_{\rm w}^2 c},
% \end{equation}
% and the radiation energy density is given by $U_{\rm sh} = 3p_{\rm sh}$. If the radiation field is in LTE, then the radiation temperature can be estimated by
% \begin{equation}
%     T_{\rm bo} = \lrb{U_{\rm sh}/a}^{1/4} \simeq 3\times10^5\mr{\,K}\, (\Lj/10^)^{1/4} {r_{\rm w}/3\times10^{14}\,\mr{cm}}^{-1/2},
% \end{equation}
% which means that the shock breakout will be bright in the soft X-ray band. The volume of the breakout region is roughly given by $4\pi \sin\theLS \thej r_{\rm w}^3\sim r_{\rm w}^3$ (for $\theLS\sim 1\rm\, rad$ and $\thej\sim 0.1\rm rad$), so the total radiated energy is given by $\sim U_{\rm sh} r_{\rm w}$. We assume that this energy is radiated away in a quasi-isotropic manner (ignore the weak beaming) on a timescale of $P_{\rm prec}$, the luminosity of the shock break is given by
% \begin{equation}
%     L_{\rm bo} \simeq 
% \end{equation}

% from the shocks are collisionless as opposed to radiation mediated and we expect bright non-thermal emission to be produced when the jet is not choked by the disk wind.

(4) We have assumed the jet isotropic luminosity $\Lj$, wind mass loss rate $\dotMw$, and the misalignment angle $\theLS$ to be constant in this paper, whereas they are all time-dependent in realistic TDEs --- we expect $\Lj$ and $\dotMw$ to evolve on a timescale that is comparable to the time since the tidal disruption and $\theLS$ to evolve on the timescale over which the jet becomes aligned with the BH spin axis. Do the precessing jets eventually break out of the wind confinement because of the decreasing fallback rate or Bardeen-Petterson alignment? As the fallback rate $\dot{M}_{\rm fb}$ drops, we expect the wind launching radius to track the spherization radius below which the disk becomes geometrically thick \citep{yuan14_ADAF_review}, which is roughly given by $\rg (\dot{M}_{\rm fb} c^2/L_{\rm Edd})$, where $L_{\rm Edd}$ is the Eddington luminosity of the BH. As the wind launching radius (denoted as $\rd$) decreases,
this reduces the critical duty cycle (which is proportional to $\rd^{0.15}$ for $s=0.8$, cf. eq. \ref{eq:breakout_condition_BZ}) only very slightly. For this reason, we think the Bardeen-Petterson alignment of the jet plays a more important role in allowing the jet to eventually break out (provided that the jet launching process is sustained). Future works should explore longer simulations with a time-dependent misalignment angle $\theLS(t)$, which may explicitly show a choked$\rightarrow$successful jet transition.

(5) Due to finite box size, our simulations do not capture the dynamics on scales larger than about $3\times10^{15}\rm\, cm$. To test the sensitivity of our results to the outer boundary conditions, we also show the $u_{r, 90}$ distribution for $150<r/\rin<200$ in Fig. \ref{fig:breakout_general_150_200} in the Appendix and confirm our qualitative conclusions presented in \S \ref{sec:HD_sim}. However, a number of interesting physical processes occur at larger scales $r\gg 3\times10^{15}\rm\, cm$, including internal shocks between adjacent jet windings and external shocks between the fast outflow and circum-nuclear medium. Future simulations should study these shocks which are expected to be collisionless and hence will produce bright non-thermal emission \citep{sironi15_collionless_shocks}.

\section{Summary}\label{sec:summary}

In most tidal disruption events, the star's angular momentum is misaligned with the BH spin. In this paper, we \textit{assume} that the accretion disk fed by the fallback stellar debris undergoes Lense-Thirring precession around the BH spin axis and explore the consequence of this hypothesis. If relativistic jets are launched from the inner disk, they will be collimated by the wind from the outer (but geometrically thick) regions of the disk and this leads to precessing jets \citep{liska18_precessing_jets}. We analytically and numerically study the propagation of a misaligned precessing jet inside the disk wind. 

Our analytic picture is one-dimensional. Along a fixed direction that the jet sweeps across, the jet is ``on'' for a duration of $t_{\rm on}$ and ``off'' for a duration of $P_{\rm prec}-t_{\rm on}$ (and during the ``off'' time the disk wind is launched along this direction), where $P_{\rm prec}$ is the precessional period. The duty cycle is given by $\xi_{\rm duty}=t_{\rm on}/P_{\rm prec}$, which is of the order $\thej\ll 1$ for a narrow jet with half opening angle $\thej$ and large misalignment angle $\theLS\sim \rm 1\, rad$ (see Fig. \ref{fig:duty_cycle}). For such an episodic jet, we calculate the timescale $t_{\rm bo}$ for the jet head to break out from the wind ahead of it and the timescale $t_{\rm cross}$ for the reverse shock to cross the jet episode. If $t_{\rm bo}>t_{\rm cross}$, the jet is choked by the wind; and otherwise the jet will break out successfully. Based on this model, we further show that, for typical jet power according to the \citet{blandford77_BZjet} mechanism, jets with large misalignment angles $\theLS\gtrsim 20^{\rm o}$ are always choked even for maximally magnetized and spinning BHs.

We then carry out three-dimensional relativistic hydrodynamic simulations of a precessing jet. The setup is that a relativistic jet with a narrow conical opening angle precesses around the BH spin axis while there is a quasi-spherical, non-relativistic wind outflowing in all directions except the jet cone. While fixing the 4-velocity on the jet axis to be $u_{\rm j}=4.0$ (for numerical reasons) and the wind 4-velocity to be $u_{\rm w}=0.1$, we explored a wide range of misalignment angles $\theLS=0.1, 0.2, 0.4, 0.8$ and jet densities $\rhojin'/\rhowin'=0.01, 0.1, 1.0$. The case $\rhojin'/\rhowin'=0.1$ is our fiducial one as it is the closest to the physical conditions in Swift J1644+57. We also consider two jet opening angles $\thej=0.1\rm\, rad$ (fiducial) and $0.2\rm\, rad$, which affects the duty cycle factor $\xi_{\rm duty}$ together with the misalignment angle.

The results of our simulations confirm that precessing jets with large misalignment angles are choked by the disk wind. Quantitatively, jets like what was in Swift J1644+57 can only break out of the wind confinement provided that the misalignment angle is sufficiently small $\theLS\lesssim 15^{\circ} \mbox{ to } 20^{\circ}$. This is roughly in agreement with our analytic results. Based on our model, for the vast majority of the jetted TDEs, the misalignment angles are so large that the jets are choked (at least at early time when the fallback rate is near the peak value). We also find our results to be in rough agreement with the heuristic arguments proposed by \citet{teboul23_precessing_jets} except that their breakout criterion overestimates the minimum jet power for successful breakout at very large misalignment angles $\theLS\gtrsim 30^{\rm o}$.

Our model can offer a reasonable answer to why only a tiny fraction of TDEs is observed to be associated with a relativistic jet pointing near our line of sight. The event rate of Swift J1644+57-like TDEs is observationally given by $10^{-2}\mbox{ to }10^{-1}\,\rm Gpc^{-3}\,yr^{-1}$ \citep{andreoni22_AT2022cmc}, which is as small as $10^{-5}\mbox{ to }10^{-4}$ of the rate of all TDEs $\mc{R}_{\rm TDE}\sim 10^3\rm\,Gpc^{-3}\,yr^{-1}$ detected in the optical and X-ray bands \citep{vanvelzen18_TDE_rate,sazonov21_erosita_TDEs,yao23_TDE_rate}.
% Even if we account for the number of off-axis events by the beaming factor of $f_{\rm b}=\theta_{\rm j}^2/2\simeq5\times10^{-3}\,\theta_{\rm j,-1}^2$, the intrinsic jetted TDE rate is ${\cal R}_{\rm jet}\sim1-10\,\rm Gpc^{-3}\,yr^{-1}$, which is only 0.1-1\% of optical and X-ray TDEs. Note that this fraction is much smaller than the fraction of AGNs with jet, $\sim10\,\%$ \citep[e.g.,][]{Padovani+2017}.
This very small fraction could be understood if a successful prompt jet breakout happens only for ``double-alignment'' events, i.e., both the stellar angular momentum and the observer's line of sight are nearly aligned with the BH spin axis to within an angle of a few times jet opening angle $\thej$. By integrating over the probability distributions of the viewing angles and misalignment angles, we estimate that the event rate of jetted TDEs satisfying the double-alignment (DA) condition is of the order
\begin{equation}\label{eq:double_alignment_probability}
    {\mc{R}_{\rm jet, DA} \over \mc{R}_{\rm jet} } \sim {\thej\theta_{\rm LS,max}^3\over 3} \sim 10^{-3} (\thej/0.1\rm\, rad)^4,
    % \mr{few}\times \thej^4 \mc{R}_{\rm jet} \sim \mr{few} \times  10^{-4} (\theta_{\rm j}/0.1)^4 \mc{R}_{\rm jet},
\end{equation}
where we have taken $\theta_{\rm LS,max}\simeq 4\thej$ based on our analytic/numerical breakout conditions and $\mc{R}_{\rm jet}$ is the rate of all intrinsically jetted TDEs. This is in good agreement with the detection rate of Swift J1644+57-like events, provided that $(\thej/0.1\rm\, rad)^4 \mc{R}_{\rm jet}/\mc{R}_{\rm TDE}\sim 10^{-2}\mbox{ to }10^{-1}$ (meaning that 1\%--10\% of all TDEs are intrinsically jetted if $\thej=0.1\rm\, rad$).
% \tm{In the last equation, do you mean that the fraction of intrinsic jetted TDE is 1-10\% of total TDEs?}
% \wl{[WL: I adopted the simpler order-of-magnitude estimate above for the double-alignment rate. When I did a more careful calculation by integrating over the probability distributions of $\theta$ and $\theLS$ for a top-hat jet, I found the fraction of cases with successful jet breakout is roughly given by $\mc{R}_{\rm jet,DA}/\mc{R}_{\rm jet} \simeq \int_0^{\theta_{\rm LS,max}} 2\pi \theLS \d \theLS \times 1/(4\pi) \times 2\pi \theLS (2\thej) \times 2/(4\pi) = \thej \theLS^3/3$, where I have discarded the retrograde orbits with anti-aligned spin and orbital angular momentum. This seems to be a little too high, $\mc{R}_{\rm jet,DA}/\mc{R}_{\rm jet}\sim 10^{-3}(\thej/0.1)^4$ if we take $\theLS=4\thej$. I guess this could be due to our analytic breakout condition being too conservative. Going back to Fig \ref{fig:breakout_general}, it is also possible that the jet might already be choked if $u_{r, 90}\simeq 3$ instead of 2. ]}
% \begin{align}
% \frac{\theta_{\rm j}^2}{2}\frac{\theta_{\rm LS}^2}{2}\simeq4\times10^{-4}\,\theta_{\rm j,-1}^4L_{\rm j,iso,48}\beta_{\rm w,-1}^{-1}\left(\frac{\dot{M}_{\rm w}}{\Msun \rm yr^{-1}}\right)^{-1}\ ,
% \end{align}
% \begin{equation}
%     {\thej \theta_{\rm LS,max}^3\over 3} \simeq 
% \end{equation}
% which is roughly consistent with the fraction of on-axis jetted TDEs.
% \note{Our interpretation of Swift J1644+57 (being double-alignment) is different from that in \citet{teboul23_precessing_jets}.}

Our model also leads to a number of testable predictions:

(1) We expect a large population of TDEs which have a successful jet breakout only at late time (long after the peak fallback rate) due to Bardeen-Petterson alignment of the accretion disk with the BH spin. These late-time jets are largely misaligned with our (random) line of sight, and their event rate can be estimated to be $10$--$10^2\rm\, Gpc^{-3}\,yr^{-1}$, which is a factor of the order $\mc{R}_{\rm jet}/\mc{R}_{\rm jet,DA}\sim 10^3(\thej/0.1)^{-1}$ higher than that of Swift J1644+57-like events \citep[0.01--0.1$\rm \,Gpc^{-3}\,yr^{-1}$,][]{andreoni22_AT2022cmc}.
% The rate of these TDEs can be inferred based on our interpretation that Swift J1644+57-like events are the fortunate double-alignment cases where (i) the stellar angular momentum is nearly aligned with the BH spin to within an angle of the order $\theLS\sim \mr{few}\, \thej$ and (ii) the observer's line of sight is nearly aligned with the BH spin to within an angle of the order $\theta\sim \mr{few}\,\thej$. If we take these two beaming factors to be of the order $1/30$ each and detection rate of Swift J1644+57-like events to be $0.01$--$0.1\rm\, Gpc^{-3}\,yr^{-1}$ \citep[e.g.,][]{andreoni22_AT2022cmc}, we infer that the rate of misaligned TDEs with late-time jet breakout to be $10$--$100\rm\, Gpc^{-3}\,yr^{-1}$.\tm{Is this the number of intrinsic jet-launching TDEs?}
% This fraction comes from our interpretation that the Swift J1644+57-like events (mostly viewed far away from the jet beaming cone by an observer on the Earth), which account for $\gtrsim 10^{-3}$ of all TDEs, are the ones where the stellar angular momentum is nearly aligned with the BH spin to within an angle of the order $\theLS\sim \thej\sim \mc{O}(0.1)$. The number of sources with highly misaligned stellar angular momenta is larger than the nearly aligned ones by a factor of the order $\thej^{-2}\sim \mc{O}(10^{2})$. In these sources, the precessing jets will initially be choked by the disk wind, but at late time when the disk is brought to alignment with the BH spin, the jets will eventually break out and then produce bright radio afterglow emission.
A possible observational support of this scenario is the late-time radio re-brightening seen in a large fraction (up to 40\%) of optically-selected TDEs \citep{cendes23_late_radio} --- the rapid rise of the radio flux a few years after the TDEs is consistent with an off-axis jet model \citep{matsumoto23_off-axis_jet_equipartition, sfaradi23_AT2018hyz_off-axis_jet}.
% \tm{I got puzzled by this statement. Why do delayed radio flares support delayed jet breakout? I agree that when a jet becomes aligned with BH spin axis, it can break out and typically its direction is different from our line of sight. In such a case, we cannot detect radio, right? (Here I assume a relativistic jet) Then the event rate is at most a few \% of optical TDEs, which is still smaller than the rate estimated by Cendes. Are you thinking of radio emission emitted by decelerated (Newtonian) jet? In this case, you statement sounds at least 40\% of optical TDEs intrinsically launch a jet. Am I correct?}
We also note that delayed disk-spin alignment makes it possible for the observer to see the X-ray emission from the inner regions of the disk at late time. This provides a possible explanation for the delayed X-ray brightening in most optically selected TDEs \citep{guolo23_TDE_Xrays}, although some other alternative explanations \citep[e.g., the accretion disk becoming more and more geometrically thin,][]{wen20_disk_thinning, thomsen22_superEdd_outflows} are also viable.

(2) We expect a small population of TDEs which show bright, spectroscopically hard, and rapidly variable X-ray emission at late time after the delayed jet breakout, provided that the observer's line of sight is nearly aligned with the BH spin axis $\theta\sim \mr{few}\,\thej$ --- these are the ``single alignment'' cases. Following the argument in the earlier paragraph, we infer the rate of these events to be roughly $1$--$10\rm\, Gpc^{-3}\,yr^{-1}$. This prediction can be tested in the near future by (i) systematic X-ray follow-up observations of known TDEs at late time and (ii) all-sky X-ray surveys such as eROSITA \citep[e.g.,][]{sazonov21_erosita_TDEs} and Einstein Probe \citep{yuan22_einstein_probe}.

% delayed jet breakout. For a line of sight close to the BH spin axis, an observer may only detect bright optical emission at early time but then bright, rapidly varying X-ray emission is emitted from the jet at late time. The rate of these sources is much higher than the detection rate of Swift J1644+57-like events by a factor of the order $\thej^{-2}\sim 10^2$ because the star's angular momentum does not have to be nearly aligned with the BH spin.

(3) Our simulations show that weakly misaligned ($\theLS\sim \mr{few}\,\thej$), precessing jets will successfully break out of the wind confinement. We expect the afterglow emission from such successful, weakly precessing jets to be rather different from the standard case of a non-precessing jet \citep[e.g.,][]{granot02_on-axis_afterglows}. The main difference is that a large amount of kinetic energy is beamed away from any given line of sight. A precessing jet occupies a solid angle of the order $\sim 4\pi \theLS \thej$, whereas an observer at a viewing angle of $\theta\simeq \theLS$ away from the BH spin axis only sees the emission from a solid angle of $\sim\! \pi \thej^2$ (provided that the jet Lorentz factor is of the order $\Gamj\sim 1/\thej$). The ratio between these two solid angles is $4\theLS/\thej$, which may be up to an order of magnitude. This means that, as the entire jet decelerates to mildly relativistic speeds, the energy contained in the observable synchrotron emitting region (that is heated by the forward shock) will gradually rise by up to an order of magnitude. This is consistent with what has been inferred from the late-time radio afterglow of Swift J1644+57 \citep[e.g.,][]{berger12_J1644_afterglow, barniolduran13_J1644_afterglow_energy}. Recently, \citet{beniamini23_J1644_afterglow} modeled the afterglow data of Swift J1644+57 with a non-precessing jet viewed slightly off-axis; in their model, the energy in the emitting region increases over time as the jet decelerates and hence an increasing fraction of the forward-shock-heated region becomes visible to the observer. According to our model, the jet in Swift J1644+57 could be weakly misaligned with the BH spin axis and our line of sight is within the angular region that the jet sweeps through. This is supported by the quasi-period oscillations in the X-ray lightcurve on timescales of a few days \citep{saxton12_J1644_QPO, lei13_J1644_misaligned} and that the duty cycle factor is of order unity.

% There is a very small population of TDEs with small misalignment angles $\theLS\sim \mr{few}\, \thej$ and small viewing angles $\theta\sim \mr{few}\,\thej$ (the ``near double alignment'' cases), and their precessing jets will break out of the wind confinement at early time. Such jets have a large duty-cycle factor of order unity, but their X-ray lightcurves may still be periodically modulated by the jet precession. The afterglow emission of these events are different from the standard on-axis afterglow of a non-precessing jet \citep[e.g.,][]{granot02_on-axis_afterglows}. This is because most of the jet energy is directed away from the observer's line of sight and we expect the energy in the observable synchrotron emitting region to increase over time until the entire jet decelerates to mildly relativistic speeds. In fact, Swift J1644+57 might be one of these cases as  and its radio afterglow indicates that the energy in the visible shock-heated region increases by an order of magnitude in the first year \citep{berger12_J1644_afterglow, barniolduran13_J1644_afterglow_energy, beniamini23_J1644_afterglow}.

(4) We expect PeV neutrinos to be produced by TDEs with choked precessing jets. The collisionless shocks in choked jets will accelerate cosmic rays to very high energies possibly up to $10^{20}\rm\, eV$ \citep{farrar14_cosmic_rays}. The cosmic ray protons will interact with thermal photons and produce charged pions which later produce neutrinoes. The mean free path of a cosmic ray proton with energy near the $\Delta$ resonance is roughly given by $\ell_{p\gamma} = (n_\gamma \sigma_{p\gamma})^{-1}\sim  10^{14}\mr{\,cm} (T/10^4\mr{\,K})^{-3}$, where $\sigma_{p\gamma}\simeq 5\times10^{-28}\rm\, cm^{-2}$ is the interaction cross-section and we have taken the photon number density to be $n_\gamma\sim 4\sigma_{\rm SB}T^4/(3 ck_{\rm B}T)$ for thermal radiation field at temperature $T$, for Stefan-Boltzmann constant $\sigma_{\rm SB}$ and Boltzmann constant $k_{\rm B}$. For typical temperatures of $T\gtrsim 10^{4}\rm\, K$ as seen in optically selected TDEs, the mean free path is much shorter than the typical jet-wind interaction radius of $r_{\rm w}\sim 3\times 10^{14}\rm\, cm$. Moreover, cosmic ray protons with energies near $10^{16}\rm\, eV$ (corresponding to Lorentz factors $\sim 10^7$) will strongly interact with thermal photons of energies near $3k_{\rm B}T\sim 10\rm\, eV$, and such interactions convert about 5\% of the proton's energy into neutrinos \citep{kelner08_p-gamma_interactions}, so we expect TDEs with choked jets to be factories of $\sim$PeV neutrinos \citep[see also][but they only considered aligned jets whose event rate is low]{wang16_PeV_neutrinos, zheng23_neutrinos_choked_jets, mukhopadhyay23_neutrinos_from_delayed_jets}. There are existing observational supports to this prediction \citep{stein21_TDE_neutrino, vanvelzen21_TDE_neutrinos, reusch22_AT2019fdr_neutrino, jiang23_TDE_neutrinos} \citep[see also][who considered other non-jetted explanations]{hayasaki19_TDE_disk_neutrinos, murase20_TDE_wind_neutrinos}.

% \note{to be continued}

\section*{Acknowledgements}
WL acknowledges the support from the Rose Hills Innovator Program. TM acknowledges supports from JSPS Overseas Research Fellowship and the Hakubi project at Kyoto University.  CDM is supported by an NSERC Discovery Grant.

\section*{Data Availability}
The data produced in this study will be shared on reasonable request to the authors.

\bibliographystyle{mnras}
\bibliography{references.bib}

\appendix
\section{Additional Figures}

\begin{figure*}
\centering
\includegraphics[width=0.75\textwidth]{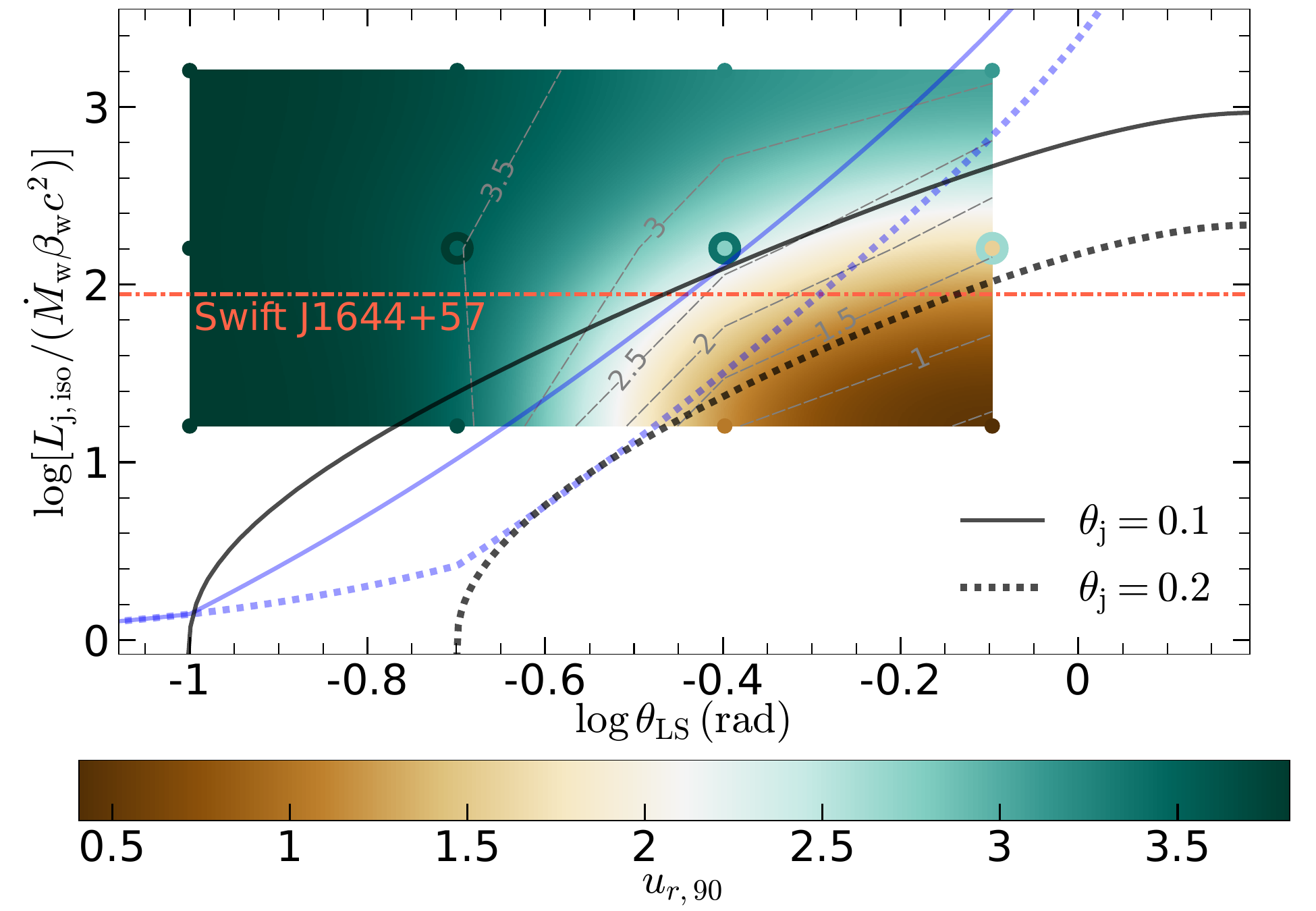}
\caption{Same as Fig. \ref{fig:breakout_general} except that $u_{r,90}$ is measured in the radial range of $150<r/\rin<200$.
  }
\label{fig:breakout_general_150_200}
\end{figure*}

\begin{figure*}
\centering
\includegraphics[width=0.45\textwidth]{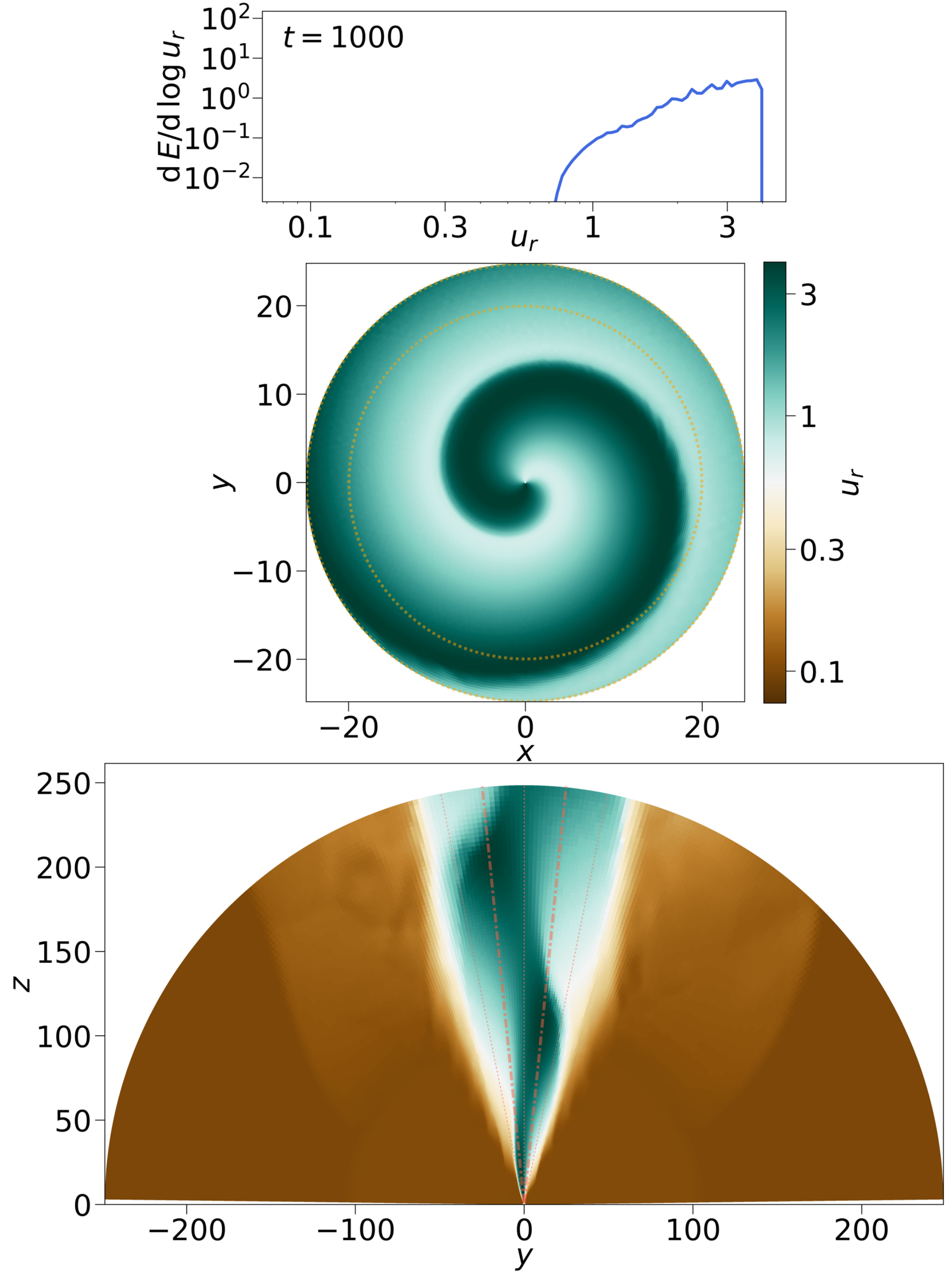}
\includegraphics[width=0.45\textwidth]{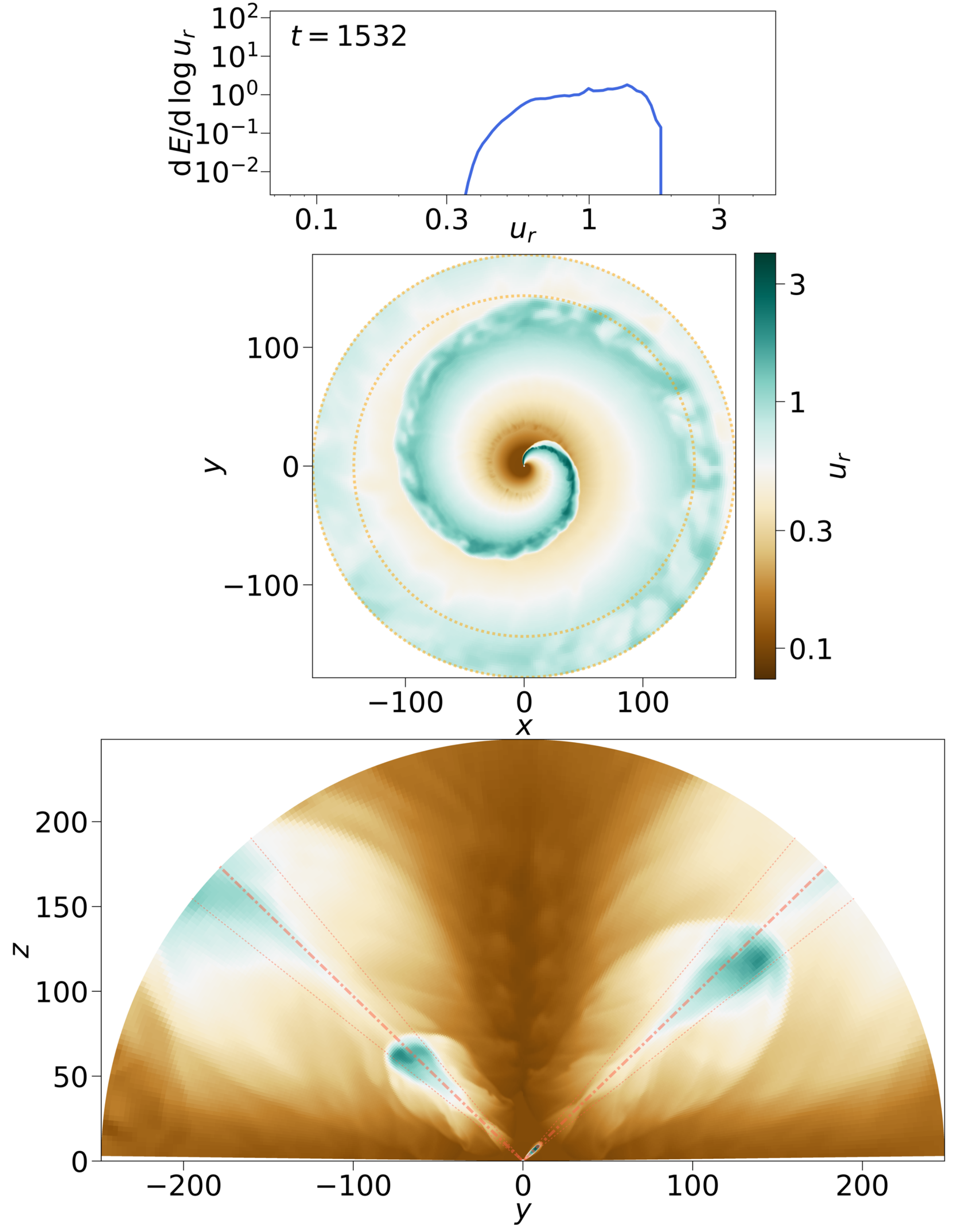}
\caption{Steady-state snapshots of the simulations of our fiducial cases for the jet power and wind mass-loss rate $\rhojin'/\rhowin'=0.1, \Omgp=0.03\rin/c, \thej=0.1\rm\, rad$. The two panels show different misalignment angles $\theLS=0.1\rm\, rad$ (left) and $0.8\rm\, rad$ (right). The upper panels show the energy distribution in radial 4-velocity space (same as in Fig. \ref{fig:dEdur}). The middle panels show the radial 4-velocity profile on a cone slice at $\theta=\theLS$ that is projected onto the x-y plane. Two orange dashed lines show $r=200$ and $r=250$. The bottom panels show the radial 4-velocity profile on a planar slice in the y-z plane. The red dash-dotted lines show the jet axis, around which two thinner red dotted lines show the jet boundaries at polar angles $\theta=\theLS\pm\thej$. The time $t$ is in units of $\rin/c$.
  }
\label{fig:rhojin0.1}
\end{figure*}

\begin{figure*}
\centering
\includegraphics[width=0.45\textwidth]{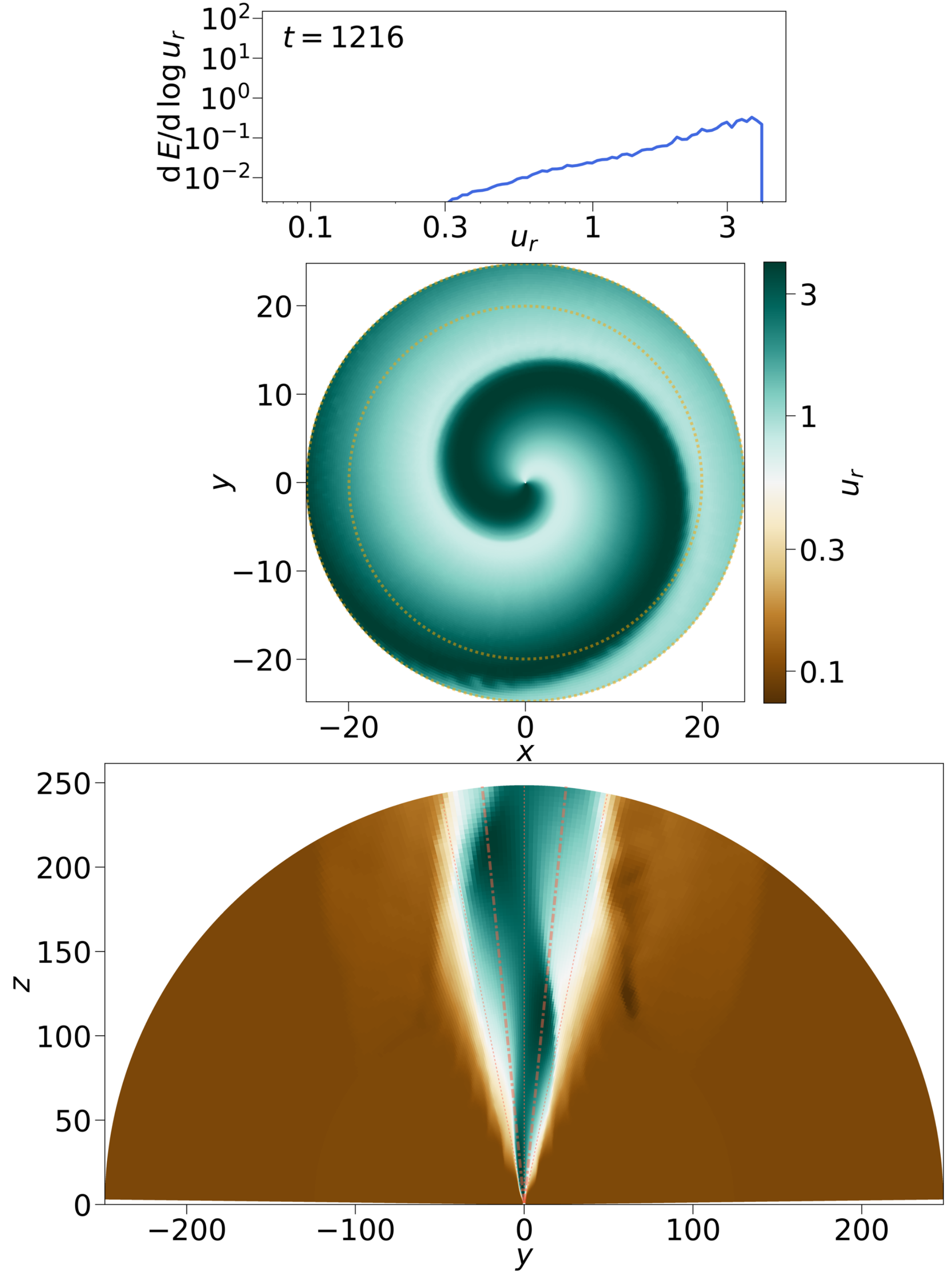}
\includegraphics[width=0.45\textwidth]{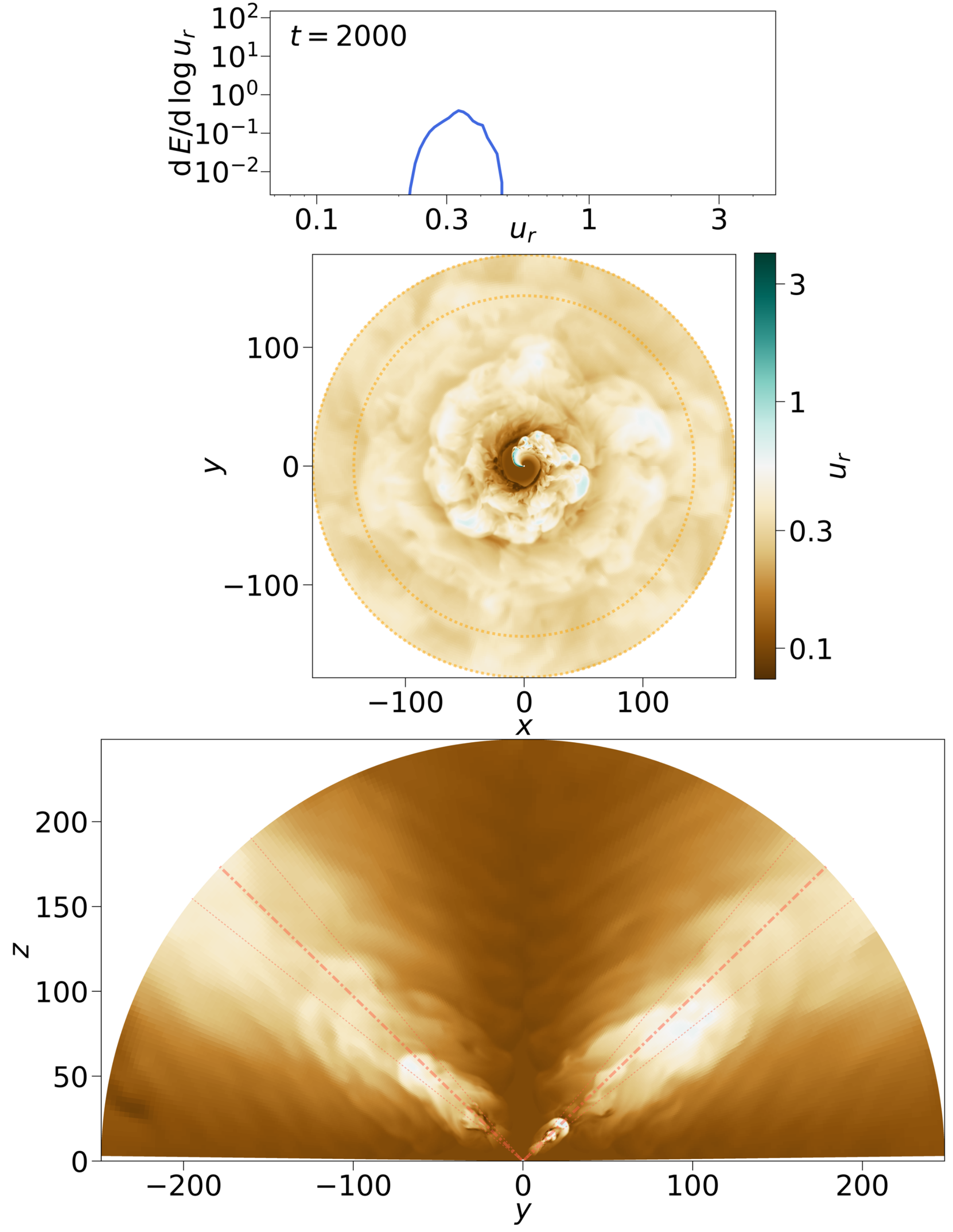}
\caption{Same as Fig. \ref{fig:rhojin0.1}, except for a much weaker jet with $\rhojin'/\rhowin'=0.01$.
  }
\label{fig:rhojin0.01}
\end{figure*}

\begin{figure*}
\centering
\includegraphics[width=0.45\textwidth]{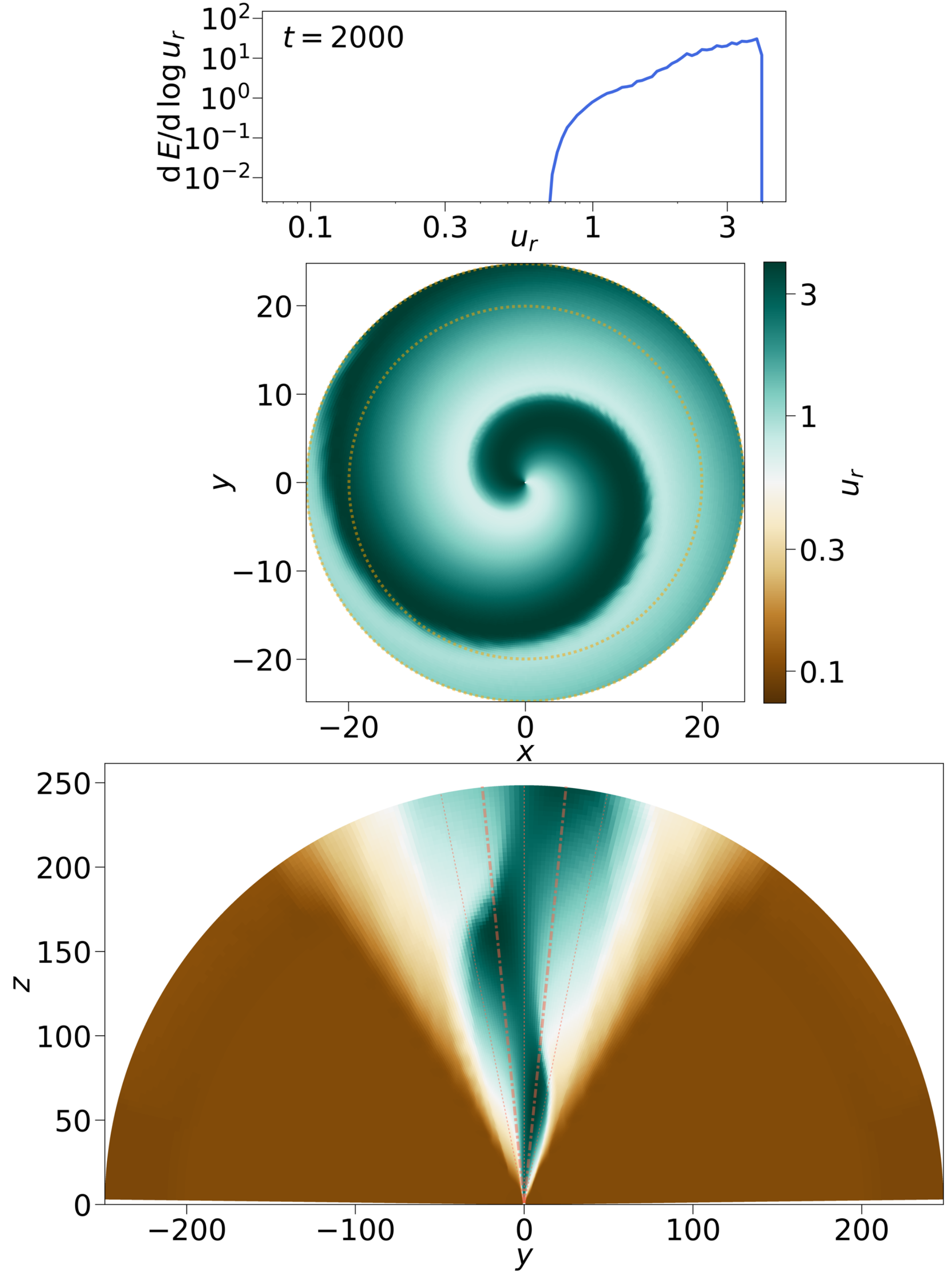}
\includegraphics[width=0.45\textwidth]{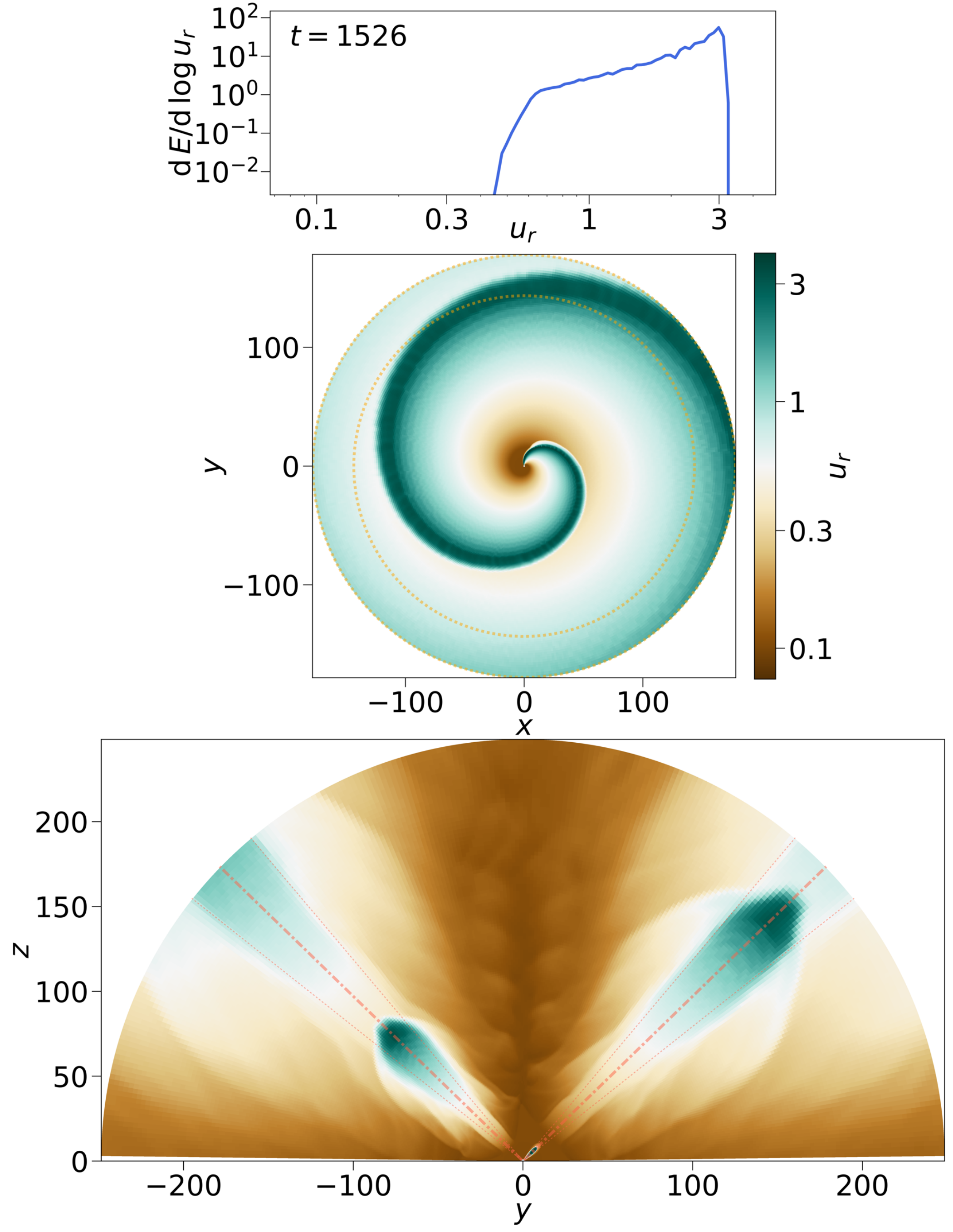}
\caption{Same as Fig. \ref{fig:rhojin0.1}, except for a much stronger jet with $\rhojin'/\rhowin'=1.0$.
  }
\label{fig:rhojin1.0}
\end{figure*}

\end{document}